\documentstyle[12pt,epsfig]{article}

\def\gsim{\mathrel{\rlap {\raise.5ex\hbox{$ > $}}
{\lower.5ex\hbox{$\sim$}}}}
\def\lsim{\mathrel{\rlap {\raise.5ex\hbox{$ < $}}
{\lower.5ex\hbox{$\sim$}}}}

\newcommand{\be}{\begin{equation}}
\newcommand{\ee}{\end{equation}}
\newcommand{\bea}{\begin{eqnarray}}

\newcommand{\eea}{\end{eqnarray}}

\def\gappeq{\mathrel{\rlap {\raise.5ex\hbox{$>$}}
{\lower.5ex\hbox{$\sim$}}}}

\def\lappeq{\mathrel{\rlap{\raise.5ex\hbox{$<$}}
{\lower.5ex\hbox{$\sim$}}}}

\begin{document}

\begin{titlepage}
\vspace{0.1in}

\begin{center}

{\Large {\bf Ternary "Quaternions"\\
 and Ternary $TU(3)$ algebra}}\\

 \vspace{0.2in}

\Large{\it  Guennady Volkov }

\begin{flushright}
PNPI--- St-Peterbourg\\
CERN--- Geneve\\
\end{flushright}

\end{center}

\vspace{0.2in}

\date{25.6.2010}
\vspace{0.2in}

\begin{abstract}

\vspace{0.2in}
To construct  ternary "quaternions" following Hamilton we must
introduce two "imaginary "units, $q_1$ and $q_2$ with propeties $q_1^n=1$ and $q_2^m=1$. The general is enough difficult, and we consider the $m=n=3$.
This case gives us the example of non-Abelian groupas was in Hamiltonian
quaternion. The Hamiltonian quaternions help us to discover the $SU(2)=S^3$ group and also the group $L(2,C)$. In ternary case we found the generalization of U(3) which we called $TU(3)$ group and also we found the the SL(3,C) group. On the matrix language we are going from binary Pauly matrices to three dimensional nine matrices which are called by nonions.
 This way was  initiated by algebraic
classification of $CY_m$-spaces for all m=3,4,...where in reflexive Newton polyhedra we found the Berger graphs which gave in the corresponding Cartan matrices
 the longest simple
roots  $B_{ii}=3,4,..$ comparing with
the case of binary algebras in which the Cartan diagonal element
is equal 2, {\it i.e. } $A_{ii}=2$.

We will discuss the following results

\begin{itemize}
\item{${}_n$ quaternization of ${R}^n $ spaces}\\
\item{Ternary "quaternion" structure structure and the  invariant surfaces}\\
\item{New geometry and non-Abelian N-ary algebras/symmetries}\\
\item{Root system of a new ternary $TU(3)$ algebra}\\
\item{N-ary Clifford algebras}
\end{itemize}

\end{abstract}

\end{titlepage}
\tableofcontents

\newpage
\section{Introduction}

The complexification of $R^n$ Euclidean spaces gave us the generalization of $U(1)$ group to the n-parameter  Abelian groups
$ U_n=\exp(\alpha_1 q+\alpha_2q^2+....+\alpha_{n-1}q^{n-1})$
\cite{LRV,LV, Volkov1}
The Hamilton procedure is going to discover non-Abelian groups \cite {Hamilton} .
This question we will discuss in our article.

 In all these approaches there were
used a wide class of simple classical Lie algebras, whose
Cartan-Killing classification  contains four infinite series
$A_r=sl(r+1)$,$B_r=so(2n+1)$, $C_r=sp(2r)$, $D_r=so(2r)$ and five
exceptional algebras ${\it G}_2$, ${\it F}_4$, ${\it E}_6$, ${\it
E}_7$, ${\it E}_8$.
  There were used some ways to study such
classification. We can remind some  of them,
 one way is through the theory of numbers and Clifford
algebras, the second is  the geometrical way,  and at last,
the third is through the theory of Cartan matrices and Dynkin diagramms.

 Before to show a new root system for ternary non-Abelian algebra (in our example of $(TU(3)$ )it is very useful to remind the theory of simple roots in binary Cartan Lie algebra. We well know how the simple roots  allows us to reconstruct all root system and, consequently,
all commutation relations in the corresponding CLA.

The finite-dimensional Lie algebra {\it g} of a compact simple Lie
group $G$ is determined by the following binary commutation relations

\begin{equation}
[T_a,T_b]_{Z_2}={\bf i} f_{abc}\,T_c,
\end{equation}
where the basis of generators $\{T_a\}$ of ${\it g}$ is assumed to
satisfy the orthonormality condition:

\begin{equation}
Tr(T_aT_b)=y\, \delta_{ab}.
\end{equation}
The constant $y$ depends on the representation chosen.

The standard way of choosing a basis for {\it g} is to define the
maximal set ${\it h}, \, [{\it h}{\it h}]=0$, of commuting
Hermitiam generators, $H_i$, (i=1,2,...,r).
\begin{equation}
 [H_i,H_j]=0, \qquad 1 \leq i,\, j \leq r.
\end{equation}
This set ${\it h}$ of $H_i$ forms the Abelian Cartan subalgebra
(CSA). The dimension $r$ is called the rank of {\it g} (or G).
Then we can extend a basis taking complex generators $E_{\vec
\alpha}$, such that
\begin{equation}
 [H_i,E_{\vec \alpha}]=\alpha_i E_{\vec \alpha}, \qquad 1 \leq i\leq r.
\end{equation}

From these commutation relations one can give the so called Cartan
decomposition of algebra {\it g} with respect to the subalgebra
${\it h}$:
\begin{equation}
{\it g}\,=\,{\it h} \oplus \sum_{\vec \alpha \in \Phi } {\it
g}_{\vec \alpha},
\end{equation}
where ${\it g}_{\vec \alpha}$ is one-dimensional vector space,
formed by step generator  $E_{\vec \alpha}$ corresponding to the
real $r$-dimensional vector $\vec \alpha$ which  is called a root.
$\Phi$ is a set of all roots.

For each $\vec \alpha$ there is one essential step operator
$E_{\vec \alpha} \in {\it g}_{\vec \alpha}$ and for $-\vec \alpha$
there exist the step operator $E_{-\vec \alpha} \in {\it g}_{-\vec
\alpha}$ and
\begin{equation}
E_{-\vec \alpha}={E_{\vec \alpha}}^{\star}.
\end{equation}

It is convenient to form a basis for r-dimensional root space
$\Phi$. It is well-known that a basis $\vec \alpha_{1},\ldots,
\vec \alpha_{r} \in \Pi \subset \Phi$ can be chosen in such a way
that for any root $\vec \alpha \in \Phi$ one can get that

\begin{equation}
\vec \alpha = \sum_{i=1}^{i=r}n_i \vec \alpha_{i},
\end{equation}
where each $n_i \in Z$ and either $n_i \leq 0$, $1 \leq i\leq r$,
or $n_i \geq 0$,  $1 \leq i\leq r$. In the former case $\vec
\alpha$ is said to be positive ($\Phi^+:\vec \alpha \in \Phi^+$)
or in the latter case is negative ($\Phi^-:\vec \alpha \in
\Phi^-$).

Such basis is basis of simple  roots.

So if such a basis is constructed one can see that for each $\vec
\alpha \in \Phi^+ \subset \Phi$, the set of the non-zero roots
$\Phi$ contains itself $-\vec \alpha \in \Phi^- \subset \Phi$,
such that

\begin{equation}
\Phi\,=\,\Phi^+ \cup \Phi^-,\qquad  \Phi^-\,=\,-\Phi^+.
\end{equation}

To complete the statement of algebra {\it g} we need to consider
$[E_{\vec \alpha},E_{\vec \beta}] $ for each pair of roots
$\vec\alpha, \vec \beta$. From the Jacobi identity one can get
\begin{equation}
 [H_i,[E_{\vec \alpha},E_{\vec \beta}]]=(\alpha_i+\beta_i)
[E_{\vec \alpha},E_{\vec \beta}].
\end{equation}
From this one can get

\begin{eqnarray}
 [E_{\vec \alpha},E_{\vec \beta}]&
=&{\it N}_{\vec \alpha, \vec \beta}E_{\vec \alpha+\vec \beta},
\qquad if \qquad \vec \alpha
+\vec \beta \qquad \in \Phi              \nonumber\\
&=&2 \frac{\vec \alpha \cdot \vec H}{<\vec \alpha,\vec \alpha>},
\qquad if \qquad \vec \alpha +\vec \beta =0,\nonumber\\
&=&0, \qquad \qquad \qquad otherwise.
\end{eqnarray}

All this choice of generators is called a Cartan-Weyl basis. For
each root $\vec \alpha$,
\begin{equation}
\{\, E_{\vec \alpha}, \qquad 2\,\frac{\vec \alpha \cdot \vec
H}{<\vec \alpha,\vec \alpha>}, \qquad E_{-\vec \alpha}\, \}
\end{equation}
form an $su(2)$ subalgebra, isomorphic to
\begin{equation}
\{I_{+}, \qquad 2I_3 , \qquad  I_{-}\, \},
\end{equation}
where

\begin{equation}
[I_{+},I_{-}]=2I_3, \qquad [I_3,I_{\pm}]=\pm I_{\pm}
\end{equation}
with
\begin{eqnarray}
I_+^{*}=I_, \qquad \qquad  I_3^{*}=I_3.
\end{eqnarray}

As consequence one can expect that the eigenvalues of
$2\,\frac{\vec \alpha \cdot \vec H}{<\vec \alpha,\vec \alpha>}$
are integral,   i.e.:
\begin{equation}
2\, \frac{< \vec \alpha, \vec \beta>}{<\vec \alpha, \vec \alpha>}
\in Z
\end{equation}
for all roots $ \vec \alpha$, $\vec \beta$.

As the examples one can consider one can consider the root systems
for $su(3)$  of rank $2$ (see $su(3)$ root system).

Now we introduce the plus-step operators:

\begin{eqnarray}
Q_1=Q_I^+=\left(
\begin{array}{ccc}
0 & 1 & 0 \\
0 & 0 & 0 \\
0 & 0 & 0 \\
\end{array}
\right)
\qquad
Q_2=Q_{II}^+=\left(
\begin{array}{ccc}
0 & 0 & 0 \\
0 & 0 & 1 \\
0 & 0 & 0 \\
\end{array}
\right)
\qquad
Q_3=Q_{III}^+=\left(
\begin{array}{ccc}
0 & 0 & 0 \\
0 & 0 & 0 \\
1 & 0 & 0 \\
\end{array}
\right)
\end{eqnarray}

and on the minus-step operators:
\begin{eqnarray}
Q_4=Q_{I}^-=\left(
\begin{array}{ccc}
0 & 0 & 0 \\
1 & 0 & 0 \\
0 & 0 & 0 \\
\end{array}
\right)
\qquad
Q_5 =Q_{II}^-=\left(
\begin{array}{ccc}
0 & 0 & 0 \\
0 & 0 & 0 \\
0 & 1 & 0 \\
\end{array}
\right)
\qquad
Q_6=Q_{III}^-=\left(
\begin{array}{ccc}
0 & 0 & 1 \\
0 & 0 & 0 \\
0 & 0 & 0 \\
\end{array}
\right)
\end{eqnarray}

We choose the following 3-diagonal
operators :

\begin{eqnarray}
H_3=Q_7=Q_{I}^0=\left(
\begin{array}{ccc}
\frac{1}{\sqrt{6}} &       0            &       0                   \\
0                  & \frac{1}{\sqrt{6}} &       0                   \\
0                  &        0           & -\sqrt{\frac{2}{3}} \\
\end{array}
\right)
\qquad
H_8=Q_8=Q_{II}^0=\left(
\begin{array}{ccc}
\frac{1}{\sqrt{2}} & 0                   & 0 \\
0                  & -\frac{1}{\sqrt{2}} & 0 \\
0                  & 0                   & 0 \\
\end{array}
\right)
\qquad
Q_0=Q_{III}^0=\left(
\begin{array}{ccc}
\frac{1}{\sqrt{3}}& 0                  & 0                 \\
0                 &  \frac{1}{\sqrt{3}}& 0                 \\
0                 & 0                  & \frac{1}{\sqrt{3}}\\
\end{array}
\right),
\end{eqnarray}

For $su(3)$ algebra the positive
roots can be chosen as
\begin{eqnarray}
\vec{\alpha}_1=(1,0),\qquad
\vec{\alpha}_2=(-\frac{1}{2},\frac{\sqrt{3}}{2}), \qquad
\vec{\alpha}_1+\vec{\alpha}_2= \vec{\alpha}_3=
(\frac{1}{2},\frac{\sqrt{3}}{2})
\end{eqnarray}
with
\begin{eqnarray}
<\vec \alpha_1,\vec \alpha_1>=<\vec \alpha_2,\vec \alpha_2>=1
\qquad  and \qquad <\vec \alpha_1 ,\vec \alpha_2>=-\frac{1}{2}.
\end{eqnarray}

\begin{eqnarray}
&&[\vec H, Q_{\pm {\vec \alpha}_1}] =\pm (\,1,0)\, Q_{\pm {\vec
\alpha}_1};\qquad \qquad [Q_{ {\vec \alpha}_1},Q_{- {\vec
\alpha}_1}] =2(1,0)\cdot \vec H;
\nonumber\\
&&[\vec H, Q_{\pm {\vec \alpha}_2}] =\pm (-\frac{1}{2},
\frac{\sqrt{3}}{2})\, Q_{\pm {\vec \alpha}_2},\qquad \,\, [Q_{
{\vec \alpha}_2},Q_{- {\vec \alpha}_2}]
=2(-\frac{1}{2},\frac{\sqrt{3}}{2})\cdot \vec H;
\nonumber\\
&&[\vec H, Q_{\pm {\vec\alpha}_3}] =\pm
(\,\frac{1}{2},\frac{\sqrt{3}}{2})\, Q_{\pm {\vec
\alpha}_3},\qquad \,\, [E_{ {\vec \alpha}_3},E_{- {\vec
\alpha}_3}] =2(\,\frac{1}{2},\frac{\sqrt{3}}{2})\cdot \vec H;
\nonumber\\
\end{eqnarray}
where $\vec H=(H_3,H_8)$. The commutation relations of step
operators can be also easily written:

\begin{eqnarray}
&&[Q_{{\vec \alpha}_1},
 Q_{{\vec \alpha}_2}]
=Q_{{\vec \alpha}_3},\qquad \qquad [Q_{{\vec \alpha}_1}, Q_{{\vec
\alpha}_3}]
=[Q_{{\vec \alpha}_2},Q_{{\vec \alpha}_3}]=0\nonumber\\
&&[Q_{ {\vec \alpha}_1}, Q_{-{\vec \alpha}_3}] =Q_{-{\vec
\alpha}_2},  \qquad \qquad [Q_{ {\vec \alpha}_2}, Q_{-{\vec
\alpha}_3}]
=Q_{-{\vec \alpha}_1}. \nonumber\\
\end{eqnarray}

For $A_2$  algebra the nonzero  roots can be also expressed
through the orthonormal basis $\{\vec e_i\}, i=1,2,3$, in which
all the roots are lying on the plane ortogonal to the vector
${\vec k}=1 \cdot {\vec e}_1+1 \cdot {\vec e}_1+1 \cdot {\vec
e}_1$, i.e.${\vec k} \cdot {\vec \alpha}=0$. Then  for this algebra the
positive roots are  the following:
\begin{eqnarray}
{\vec\alpha}_1={\vec e}_1-{\vec e}_2, \qquad {\vec\alpha}_2={\vec
e}_2-{\vec e}_3, \qquad {\vec\alpha}_3={\vec e}_1-{\vec e}_3.
\end{eqnarray}

This basis can be practically used in general case
to give the complete   list of simple finite dimensional Lie algebras

\begin{eqnarray}
\begin{array}{ccccc}
SU(n)    &  \pm (e_i-e_j)   & 1 \leq i \leq j \leq n & 0  & (n-1)  \\
SO(2n)   &  \pm e_i\pm e_j  & 1 \leq i \leq j \leq n & 0  &  n     \\
SO(2n+1) &  \pm e_i\pm e_j  & 1 \leq i \leq j \leq n & 0  &  n     \\
         &  \pm e_i         & 1 \leq i \leq n        &    &        \\
Sp(n)    &  \pm e_i\pm e_j  & 1 \leq i \leq j \leq n & 0  &  n     \\
         &  \pm 2e_i         & 1 \leq i \leq n        &    &        \\
\end{array}
\end{eqnarray}

Since $su(n)$ is the Lie algebra of traceless $n \times n$ anti-Hermitian matrices
there being $(n-1)$ linear independent diagonal matrices.
Let $h_{kl}= (e_{kk}-e_{k+1,k+1})$, $k=1,...,n-1$ be the choice of the
diagonal matrices,  and let $e_{pq}$
for $p,q=1,...,n$ , $p<q$ be the remaining the basis elements:

\begin{eqnarray}
(e_{pq})_{ks}=\delta_{kp} \delta_{sq}.
\end{eqnarray}

It is easily see that the simple finite-dimensional algebra ${\it G}$
can be encoded in  the $r \times r$  Cartan matrix
\begin{equation}
A_{ij}=\frac{<\alpha_i,\alpha_j>}{<\alpha_i,\alpha_i>}, \,\,\, 1\leq i,j \leq r,
\end{equation}

with simple roots, $\alpha_i$, which  generally  obeys  to the following rules:

\begin{eqnarray}
{ A}_{ii}&=&2 \nonumber\\
{A}_{ij}& \leq & 0\nonumber\\
{ A}_{ij}=0 &\mapsto & { A}_{ji}=0 \nonumber\\
{ A}_{ij} &\in& {Z} \qquad{= 0, 1,2,3}  \nonumber\\
Det { A} &>&0.
\end{eqnarray}

The rank of ${A}$ is equal to $r$.
\begin{eqnarray}
&&{A_r    : Det ({ A})= (r+1)},                    \nonumber\\
&&{D_r : Det ({ A})= 4    },                    \nonumber\\
&&{B_{r}  : Det ({ A})= 2    },                    \nonumber\\
&&{C_{r}  : Det ({A})= 2    },                    \nonumber\\
&&{F_4  : Det ({ A})= 1    },                    \nonumber\\
&&{G_{2}  : Det ({ A})= 1    },                    \nonumber\\
&&{E_{r}  : Det ({ A})= 9-r  }, \qquad  r=6,7,8.   \nonumber\\
\end{eqnarray}

Also using theory of the simple roots and Cartan matrices the list of
simple killing-Cartan-Lie algebras can be encoded in the Dynkin diagram.

The Dynkin diagram
of $g$ is the graph with nodes labeled $1\ldots, r$ in a bijective
correspondence with the set of the simple roots, such that nodes
$i,j$ with $i \neq j$ are joined by   $n_{ij}$ lines, where
$n_{ij}={A}_{ij} { A}_{ji}, i \neq j$.

One can easily get that $A_{ij} A_{ji}=0,1,2,3$.. Its
diagonal elements are equal 2 and its off-diagonal elements are
all negative integers or zero. The information in $A$ is codded
into Dynkin diagram which is built as follows: it consists of the
points for each simple root $\vec \alpha_{i}$ with points $\vec
\alpha_{i}$ and $\vec \alpha_{j}$ being joined by $A_{ij}A_{ji}$
lines, with arrow pointing from $\vec \alpha_{j}$ to $\vec
\alpha_{i}$ if $<\vec \alpha_{j},\vec \alpha_{j}>><\vec
\alpha_{i},\vec \alpha_{i}>$.

Obviously, that $\hat {A}_{ij}= { A}_{ij}$
for $1 \leq i,j \leq r$,
and   $\hat { A}_{00}=2$.
For generalized Cartan matrix there are two  unique vectors $ a$ and
${ a}^{\vee}$ with positive integer components $(a_0, \ldots , a_r)$
and $(a^{\vee}_0, \ldots , a^{\vee}_r)$ with their greatest
common divisor equal one, such that
\begin{eqnarray}
\sum_{i=0}^{r} a_i \hat{ A}_{ji}=0, \qquad \qquad
\sum_{i=0}^{r} \hat{A}_{ij} a^{\vee}_j =0.
\end{eqnarray}
\label{Coxeter}
The numbers, $ a_i$ and $ a_i^{\vee}$ are called Coxeter and
dual Coxeter labels.
Sums of the Coxeter and dual Coxeter labels are called by Coxeter $h$ and
dual Coxeter numbers $h^{\vee}$. For symmetric generalized Cartan matrix
the both Coxeter labels and numbers coincide. The components $a_i$,
with $i\neq 0$
are just the components of the highest root of Cartan-Lie algebra.
The Dynkin diagram
for Cartan-Lie algebra can be get from generalized  Dynkin diagram of
affine algebra
by removing one zero node. The generalized Cartan matrices and
generalized
Dynkin diagrams allow one-to-one to determine affine Kac-Moody algebras .

\newpage
\section{The geometry of ternary generalization of quaternions}

Let consider the following construction
\begin{eqnarray}
Q=z_0+z_1q_s+z_2q_s^2,
\end{eqnarray}
where
\begin{eqnarray}
z_0&=&y_0+qy_1+q^2y_2, \nonumber\\
z_1&=&y_3+qy_4+q^2y_5, \nonumber\\
z_2&=&y_6+qx_7+q^2y_8  \nonumber\\
\end{eqnarray}
are the  ternary complex numbers and $q_s$ is the new 'imaginary"
ternary unit with condition

\begin{eqnarray}
q_s^3=1
\end{eqnarray}
and
\begin{eqnarray}
q_sq=jqq_s.
\end{eqnarray}
Then one can see
\begin{eqnarray}
Q=y_0+qy_1+q^2y_2+y_3q_s+y_4qq_s+y_5q^2q_s+y_6q_s^2+y_7qq_s^2+y_8q^2q_s^2
\end{eqnarray}

We will accept the following notations:

\begin{eqnarray}
&&q=q_1,\qquad q_s=q_2, \qquad q^2q_s^2=q_3,\qquad (1) \nonumber\\
&&q^2=q_4,\qquad q_s^2=q_5, \qquad qq_s=q_6,\qquad (2) \nonumber\\
&&qq_s^2=q_7,\qquad q^2q_s=q_8, \qquad 1=q_0. \qquad (0)\nonumber\\
\end{eqnarray}
Respectively we change the notations of coordinates:

\begin{eqnarray}
&&y_1=x_1,\qquad y_3=x_2, \qquad y_8=x_3,\qquad (1) \nonumber\\
&&y_2=x_4,\qquad y_6=x_5, \qquad y_4=x_6,\qquad (2) \nonumber\\
\end{eqnarray}

In the new notations we have got  the following expression:
\begin{eqnarray}
Q
&=&(x_0+x_7q_1q^2+x_8q_1^2q_2)+(x_1q_1+x_2q_2+x_3q_1^2q_2^2)+(x_4q_1^2+x_5q_5^2+x_6q_1q_2)\nonumber\\
&\equiv &z_0(x_0,x_7,x_8)+z_1(x_1,x_2,x_3)+z_2(x_4,x_5,x_6),    \nonumber\\
\end{eqnarray}

where
\begin{eqnarray}
z_0(a,b,c)&=&a+bq_1q_2^2+cq_1^2q_2 \nonumber\\
z_1(a,b,c)&=&aq_1+bq_2+cq_1^2q_2^2 \nonumber\\
z_2(a,b,c)&=&aq_1^2+bq_2^2+cq_1q_2 \nonumber\\
\end{eqnarray}

and
\begin{eqnarray}
\{a,b,c\}=\{x_0,x_7,x_8\},\,\{x_1,x_2,x_3\},\{x_4,x_5,x_6\},
\end{eqnarray}
with all possible permutations of triples.

It is easily to check:

\begin{eqnarray}
\begin{array}{|c| c||c|c|c|}
\hline
-    & -        & \{x_0,x_7,x_8\}               & \{x_1,x_2,x_3 \}
&   \{x_4,x_5,x_6 \}   \\\hline
     & 1        & x_0+x_7q_1q_2^2 +x_8q_1^2q_2  & x_1q_1 +x_2q_2 +x_3q_1^2q_2^2
&  x_4q_1^2+x_5q_2^2+x_6q_1q_2 \\
TCl_0&q_1^2q_2  &j x_0q_1q_2^2+j^2x_7q_1^2q_2 +x_8  &j^2 x_1q_1^2q_2^2 +jx_2q_1 +x_3q_2
&  x_4q_2^2+jx_5q_1q_2+j^2x_6q_1^2 \\
     &q_1q_2^2  &j x_0q_1^2q_2+x_7 +j^2x_8q_1q_2^2  & x_1q_2 +jx_2q_1^2q_2^q +j^2x_3q_1
& j^2 x_4q_1q_2+jx_5q_1^2+x_6q_2^2 \\\hline
     &q_1       & x_0q_1^2+x_7q_2^2 +x_8q_1q_2  & x_1 +x_2q_1^2q_2 +x_3q_1q_2^2
&  x_4q_1+x_5q_1^2q_2^2+x_6q_2 \\
TCl_1&q_2       & x_0q_2^2+jx_7q_1q_2 +j^2x_8q_1^2  & jx_1q_1q_2^2 +x_2 +j^2x_3q_1^2q_2
&  j^2x_4q_1^2q_2^2+x_5q_2+jx_6q_1 \\
     &q_1^2q_2^2& j^2 x_0q_1q_2+j x_7q_1^2 +x_8q_2^2  & jx_1q_1^2q_2 +j^2x_2q_1q_2^q +x_3
&  x_4q_2+j^2x_5q_1+jx_6q_1^2q_2^2 \\\hline
     &q_1^2     & x_0q_1+x_7q_1^2q_2^2+x_8q_2& x_1q_1^2 +x_2q_1q_2 +x_3q_2^2
&  x_4+x_5q_1q_2^2+x_6q_1^2q_2 \\
TCl_2&q_2^2     & x_0q_2+j^2x_7q_1 +jx_8q_1^2q_2^2  & j^2x_1q_1q_2+x_2q_2^2 +jx_3q_1^2
 &  jx_4q_1^2q_2+x_5+j^2x_6q_1q_2^2 \\
     &q_1q_2    & j^2x_0q_1^2q_2^2+x_7q_2 +jx_8q_1  & x_1q_2^2 +j^2x_2q_1^2 +jx_3q_1q_2
&  jx_4q_1q_2^2+j^2x_5q_1^2q_2+x_6\\\hline
\end{array}
\end{eqnarray}

\begin{eqnarray}
Q&=&[z_0(x_0,x_7,x_8)+z_1(x_1,x_2,x_3)+z_2(x_4,x_5,x_6)]\nonumber\\
&=&q_1q_2^2[z_0(x_7,x_8,x_0)+z_1(x_3,x_1,x_2)+z_2(x_5,x_6,x_4)]\nonumber\\
&=&q_1^2q_2[z_0(x_8,x_0,x_7)+z_1(x_2,x_3,x_1)+z_2(x_6,x_4,x_5)]\nonumber\\
\end{eqnarray}

\begin{eqnarray}
Q&=&q_1[z_0(x_1,x_3,x_2)+z_1(x_4,x_6,x_5)+z_2(x_0,x_7,x_8)]\nonumber\\
&=&q_2[z_0(x_2,x_3,x_1)+z_1(x_5,x_4,x_6)+z_2(x_0,x_8,x_7)]\nonumber\\
&=&q_1^2q_2^2[z_0(x_3,x_2,x_1)+z_1(x_6,x_4,x_5)+z_2(x_0,x_7,x_8)]\nonumber\\
\end{eqnarray}

\begin{eqnarray}
Q&=&q_1^2[z_0(x_4,x_5,x_6)+z_1(x_0,x_8,x_7)+z_2(x_1,x_3,x_2)]\nonumber\\
&=&q_2^2[z_0(x_5,x_6,x_4)+z_1(x_7,x_0,x_8)+z_2(x_3,x_2,x_1)]\nonumber\\
&=&q_1q_2[z_0(x_6,x_4,x_5)+z_1(x_8,x_7,x_0)+z_2(x_2,x_1,x_3)]\nonumber\\
\end{eqnarray}

Now we can rewrite the expression for $Q$, $\tilde Q$, and $\tilde {\tilde Q}$ in the following way:
\begin{eqnarray}
Q&=&z_2(x_4,x_5,x_6)+q_1z_2(x_0,x_7,x_8)+q_1^2z_2(x_1,x_3,x_2)\nonumber\\
\tilde Q&=&{\tilde z}_2(x_4,x_5,x_6)+jq_1 {\tilde z}_2(x_0,x_7,x_8)+j^2q_1^2{\tilde z}_2(x_1,x_3,x_2)
\nonumber\\
\tilde{\tilde Q}&=&\tilde{\tilde z}_2(x_4,x_5,x_6)+j^2q_1\tilde{\tilde z}_2(x_0,x_7,x_8)
+jq_1^2\tilde{\tilde z}_2(x_1,x_3,x_2)\nonumber\\
\end{eqnarray}
where we accept that
\begin{eqnarray}
&&{\tilde q}_1=j q_1,\qquad {\tilde q}_2= j q_2, \nonumber\\
&&\tilde {\tilde q}_1=j^2 q_1,\qquad\tilde {\tilde q}_2. \nonumber\\
\end{eqnarray}

We would like to calculate the product $Q \tilde Q \tilde {\tilde Q}$ what in general contains itself
$9 \times 9 \times 9=729$ terms, {\it i.e.}

\begin{equation}
Q \times \tilde Q \times \tilde {\tilde Q}= A_0(x_0,...,x_8) q_0+A_1(x_0,...,x_8) q_1
+\ldots + A_8(x_0,...,x_8) q_8
\end{equation}

In general in this product one can meet inside $A_p$  $( p=0,1,...,8)$
the following term structures:

\begin{eqnarray}
&& x_p^3,\qquad p=0,1,...,8 \nonumber\\
&& x_p^2x_k,\qquad p\neq r \nonumber\\
&&x_px_kx_l, \qquad p\neq k\neq l. \nonumber\\
\end{eqnarray}
For this expansion one can easily see that
\begin{eqnarray}
729=9\,( x_p^3-terms) \,+\, 72 \times 3 \, (x_p^2x_k -terms) + 84 \times 6\, (x_px_kx_l- terms).
\end{eqnarray}

Note that we would like to save just terms proportional to $q_0=1$ - terms, {\it i.e.}
to find the $A_0$ magnitude. All others must be equal to zero.
Since $q_p^3=1$ for $p=0,1,...,8$ $A_0$ contains the first nine pure cubic terms.

We can see how in the product $Q \tilde Q \tilde {\tilde Q}$ vanish the terms
$72 \times 3 \, (x_p^2x_k -terms $.
From the expression

respectively. Now one can see that all terms disappear.
In this product we can find the nonvanishing terms proportional $q_0=1$:
\begin{eqnarray}
&& x_0^3+x_7^3+x_8^3-3x_0x_7x_8 , \nonumber\\
&& x_1^3+x_2^3+x_3^3-3x_1x_2x_3 , \nonumber\\
&& x_0^4+x_5^3+x_6^3-3x_4x_5x_6 , \nonumber\\
\end{eqnarray}
where we took into account that
\begin{eqnarray}
&&q_0q_7q_8 \sim 1 \nonumber\\
&&q_1q_2q_3 \sim 1 \nonumber\\
&&q_4q_5q_6 \sim 1. \nonumber\\
\end{eqnarray}
Also, one can also find the other combination proportional to $1$:

\begin{eqnarray}
&&q_0(q_1q_4+q_2q_5+q_3q_6)  \sim 1 \nonumber\\
&&q_7(q_1q_5+q_2q_6+q_3q_1) \sim 1 \nonumber\\
&&q_8(q_1q_6+q_2q_4+q_3q_5) \sim 1. \nonumber\\
\end{eqnarray}
So, we have got in the triple product the
\begin{eqnarray}
 9\, ( \,x_p^3\, )\, + \, 12 \times 6 \,(\, x_px_kx_l\,)\, = \,81 \,(\,terms\,).
\end{eqnarray}

The $ 729-81=72 \times 3+ 72 \times 6=648  $ terms are vanished.

Thus, we expect to get the equation for the unit ternary ``quaternion'' surface  in the following form:

\begin{eqnarray}
&&x_0^3+x_7^3+x_8^3 -3x_0x_7x_8  \nonumber\\
&&x_1^3+x_2^3+x_3^3 -3x_1x_2x_3  \nonumber\\
&&x_0^4+x_5^3+x_6^3 -3x_4x_5x_6  \nonumber\\
&&-3x_0(x_1x_4+x_2x_5+x_3x_6)    \nonumber\\
&&-3x_7(x_1x_5+x_2x_6+x_3x_4)    \nonumber\\
&&-3x_8(x_1x_6+x_2x_4+x_3x_5)=1  \nonumber\\
\end{eqnarray}

In this product we can find the nonvanishing terms proportional $q_0=1$:
\begin{eqnarray}
&& x_0^3+x_7^3+x_8^3-3x_0x_7x_8 , \nonumber\\
&& x_1^3+x_2^3+x_3^3-3x_1x_2x_3 , \nonumber\\
&& x_0^4+x_5^3+x_6^3-3x_4x_5x_6 , \nonumber\\
\end{eqnarray}
where we took into account that
\begin{eqnarray}
&&q_0q_7q_8 \sim 1 \nonumber\\
&&q_1q_2q_3 \sim 1 \nonumber\\
&&q_4q_5q_6 \sim 1. \nonumber\\
\end{eqnarray}
Also, one can also find the other combination proportional to $1$:

\begin{eqnarray}
&&q_0(q_1q_4+q_2q_5+q_3q_6)  \sim 1 \nonumber\\
&&q_7(q_1q_5+q_2q_6+q_3q_1) \sim 1 \nonumber\\
&&q_8(q_1q_6+q_2q_4+q_3q_5) \sim 1. \nonumber\\
\end{eqnarray}

\begin{eqnarray}
Q
&=&(x_0 +     x_7q_7 +     x_8q_8) +     (x_1q_1+x_2q_2+x_3q_3)  +    (x_4q_4+x_5q_5+x_6q_6) \nonumber\\
\tilde {Q}
&=&(x_0 + j   x_7q_7 + j^2 x_8q_8) + j   (x_1q_1+x_2q_2+x_3q_3)  + j^2(x_4q_4+x_5q_5+x_6q_6) \nonumber\\
\tilde {\tilde {Q}}
&=&(x_0 + j^2 x_7q_7 + j   x_8q_8) + j^2 (x_1q_1+x_2q_2+x_3q_3)  + j  (x_4q_4+x_5q_5+x_6q_6) \nonumber\\
\end{eqnarray}

\begin{eqnarray}
Q_1= q\left (
\begin{array}{ccc}
0 & 1 & 0\\
0 & 0 & 1\\
1 & 0 & 0\\
\end{array}
\right), \
\end{eqnarray}

\begin{eqnarray}
 Q_2= q^2\left (
\begin{array}{ccc}
0   & 1 & 0 \\
0   & 0 & j  \\
j^2 & 0 & 0\\
\end{array}
\right),
\end{eqnarray}

\begin{eqnarray}
 Q_3= \left (
\begin{array}{ccc}
0 & 1 & 0    \\
0 & 0 & j^2  \\
j & 0 & 0    \\
\end{array}
\right),
\nonumber\\
\end{eqnarray}

\begin{eqnarray}
Q_4= Q_1^2=q^2\left (
\begin{array}{ccc}
0 & 0 & 1    \\
1 & 0 & 0  \\
0 & 1 & 0    \\
\end{array}
\right),
\end{eqnarray}

\begin{eqnarray}
Q_5=Q_2^2=q \left (
\begin{array}{ccc}
0 & 0 & j    \\
1 & 0 & 0  \\
0 & j^2 & 0    \\
\end{array}
\right),
\end{eqnarray}

\begin{eqnarray}
Q_6=Q_3^2= \left (
\begin{array}{ccc}
0 & 0 & j^2    \\
1 & 0 & 0  \\
0 & j & 0    \\
\end{array}
\right).
\end{eqnarray}

\begin{eqnarray}
Q_7= \left (
\begin{array}{ccc}
j & 0 & 0    \\
0 & j^2 & 0  \\
0 & 0 & 1    \\
\end{array}
\right).
\end{eqnarray}

\begin{eqnarray}
Q_8= \left (
\begin{array}{ccc}
j^2 & 0 & 0    \\
0   & j & 0  \\
0   & 0 & 1    \\
\end{array}
\right).
\end{eqnarray}

\begin{eqnarray}
Q_0= \left (
\begin{array}{ccc}
1 & 0 &0    \\
0 & 1 & 0  \\
0 & 0 & 1    \\
\end{array}
\right).
\end{eqnarray}

\begin{eqnarray}
&&Q_1Q_2=j^2Q_6, \, Q_2Q_3=j^2q^2Q_4,\, Q_3Q_1=j^2qQ_5 \nonumber\\
&&Q_2Q_1=jQ_6, \, Q_3Q_2 =jq^2Q_4,\, Q_1Q_3=jqQ_5 \nonumber\\
\end{eqnarray}

\begin{eqnarray}
&&Q_4Q_5=j^2Q_3, \, Q_5Q_6=j^2q Q_1,\, Q_6Q_4=j^2q^2Q_2 \nonumber\\
&&Q_5Q_4=jQ_3, \, Q_6Q_5 =jqQ_1,\, Q_4Q_6=jq^2Q_2 \nonumber\\
\end{eqnarray}

\begin{eqnarray}
Q_1Q_5= q^2j\left (
\begin{array}{ccc}
j^2 & 0  & 0    \\
0   & j  & 0  \\
0   & 0  & 1    \\
\end{array}
\right), \qquad Q_5Q_1= q^2j^2\left (
\begin{array}{ccc}
j^2 & 0  & 0    \\
0   & j  & 0  \\
0   & 0  & 1    \\
\end{array}
\right)
\end{eqnarray}

\begin{eqnarray}
Q_1Q_6= qj^2\left (
\begin{array}{ccc}
j   & 0    & 0    \\
0   & j^2  & 0  \\
0   & 0    & 1    \\
\end{array}
\right), \qquad Q_6Q_1= qj\left (
\begin{array}{ccc}
j   & 0    & 0    \\
0   & j^2  & 0  \\
0   & 0  & 1    \\
\end{array}
\right)
\end{eqnarray}

\begin{eqnarray}
Q_2Q_4= qj^2\left (
\begin{array}{ccc}
j   & 0    & 0    \\
0   & j^2  & 0  \\
0   & 0    & 1    \\
\end{array}
\right), \qquad
Q_4Q_2= qj\left (
\begin{array}{ccc}
j   & 0    & 0    \\
0   & j^2  & 0  \\
0   & 0  & 1    \\
\end{array}
\right)
\end{eqnarray}

\begin{eqnarray}
Q_2Q_6= q^2j \left (
\begin{array}{ccc}
j^2   & 0    & 0    \\
0   & j  & 0  \\
0   & 0    & 1    \\
\end{array}
\right), \qquad
Q_6Q_2= q^2j^2\left (
\begin{array}{ccc}
j^2   & 0    & 0    \\
0   & j  & 0  \\
0   & 0  & 1    \\
\end{array}
\right)
\end{eqnarray}

\begin{eqnarray}
Q_3Q_4= q^2j\left (
\begin{array}{ccc}
j^2   & 0    & 0    \\
0   & j  & 0  \\
0   & 0    & 1    \\
\end{array}
\right), \qquad
 Q_4Q_3= q^2j^2\left (
\begin{array}{ccc}
j^2   & 0    & 0    \\
0   & j  & 0  \\
0   & 0  & 1    \\
\end{array}
\right)
\end{eqnarray}

\begin{eqnarray}
Q_3Q_5= qj^2\left (
\begin{array}{ccc}
j   & 0    & 0    \\
0   & j^2  & 0  \\
0   & 0    & 1    \\
\end{array}
\right), \qquad
 Q_5Q_3= qj\left (
\begin{array}{ccc}
j  & 0    & 0    \\
0   & j^2  & 0  \\
0   & 0  & 1    \\
\end{array}
\right)
\end{eqnarray}

\begin{eqnarray}
&&Q_1Q_4=Q_0, \, Q_1Q_5= q^2jQ_8,\, Q_1Q_6=qj^2Q_7 \nonumber\\
&&Q_4Q_1=Q_0, \, Q_5Q_1 =q^2j^2Q_8,\, Q_6Q_1=jqQ_7 \nonumber\\
\end{eqnarray}

\begin{eqnarray}
&&Q_2Q_4=qj^2Q_7, \, Q_2Q_5=Q_0,\, Q_2Q_6=j^2q^2Q_7 \nonumber\\
&&Q_4Q_2=qjQ_7, \, Q_5Q_2=Q_0,\, Q_6Q_2=jq^2Q_7 \nonumber\\
\end{eqnarray}

\begin{eqnarray}
&&Q_3Q_4=q^2jQ_8,  \, Q_3Q_5=qj^2Q_7,\, Q_3Q_6=Q_0 \nonumber\\
&&Q_4Q_3=q^2j^2jQ_8, \, Q_5Q_3=qjQ_7,\, Q_6Q_3=Q_0 \nonumber\\
\end{eqnarray}

 The  ternary conjugation include two operations:
\begin{itemize}
\item{1. $ \tilde q =j q$;}\\
\item{2.  $\{1 \rightarrow 2, 2 \rightarrow 3, 3 \rightarrow 1 \}.$}\\
\end{itemize}

Let us check the second operation. For this consider two
 $3 \times 3$ matrices:
\begin{eqnarray}
{A}= \left (
\begin{array}{ccc}
a_1 & b_1 & c_1\\
c_2 & a_2 & b_2\\
b_3 & c_3 & a_3\\
\end{array}
\right), \qquad and \qquad  {B}= \left (
\begin{array}{ccc}
u_1 & v_1 & w_1\\
w_2 & u_2 & v_2\\
v_3 & w_3 & u_3\\
\end{array}
\right )
\end{eqnarray}
Then
\begin{eqnarray}
\tilde {A}= \left (
\begin{array}{ccc}
a_3 & b_3 & c_3\\
c_1 & a_1 & b_1\\
b_2 & c_2 & a_2\\
\end{array}
\right ), \qquad and \qquad
 \tilde {B}= \left (
\begin{array}{ccc}
u_3 & v_3 & w_3\\
w_1 & u_1 & v_1\\
v_2 & w_2 & u_2\\
\end{array}
\right )
\end{eqnarray}
respectively.

Take the product of these two matrices in both cases:
\begin{eqnarray}
C={A \cdot B}= \left (
\begin{array}{ccc}
a_1u_1+b_1w_2+c_1v_3 & a_1v_1+b_1u_2+c_1w_3 & a_1w_1+b_1v_2+c_1u_3\\
c_2u_1+a_2w_2+b_2v_3 & c_2v_1+a_2u_2+b_2w_3 & c_2w_1+a_2v_2+b_2u_3\\
b_3u_1+c_3w_2+a_3v_3 & b_3v_1+c_3u_2+a_3w_3 & b_3w_1+c_3v_2+a_3u_3\\
\end{array}
\right),
\end{eqnarray}

\begin{eqnarray}
{\tilde A \cdot \tilde B}= \left (
\begin{array}{ccc}
b_3w_1+c_3v_2+a_3u_3 & b_3u_1+c_3w_2+a_3v_3 & b_3v_1+c_3u_2+a_3w_3\\
a_1w_1+b_1v_2+c_1u_3 & a_1u_1+b_1w_2+c_1v_3 & a_1v_1+b_1u_2+c_1w_3\\
c_2w_1+a_2v_2+b_2u_3 & c_2u_1+a_2w_2+b_2v_3 & c_2v_1+a_2u_2+b_2w_3\\
\end{array}
\right),
\end{eqnarray}

Compare the last expression with the expressio0n of $\tilde C$
one can see that:

\begin{eqnarray}
\tilde {(A\cdot B)}={\tilde A \cdot \tilde B}.
\end{eqnarray}

\begin{eqnarray}
\tilde{Q}_1= jq\left (
\begin{array}{ccc}
0 & 1 & 0\\
0 & 0 & 1\\
1 & 0 & 0\\
\end{array}
\right)=j Q_1,
\end{eqnarray}

\begin{eqnarray}
 {\tilde Q}_2= j^2 q^2\left (
\begin{array}{ccc}
0   & j^2 & 0 \\
0   & 0 & 1  \\
j & 0 & 0\\
\end{array}
\right)=jQ_2,
\end{eqnarray}

\begin{eqnarray}
{\tilde Q}_3= \left (
\begin{array}{ccc}
0 & j & 0    \\
0 & 0 & 1  \\
j^2 & 0 & 0    \\
\end{array}
\right)=jQ_3,
\end{eqnarray}

\begin{eqnarray}
{\tilde Q}_4=j^2q^2\left (
\begin{array}{ccc}
0 & 0 & 1    \\
1 & 0 & 0  \\
0 & 1 & 0    \\
\end{array}
\right)=j^2Q_4,
 \end{eqnarray}

\begin{eqnarray}
{\tilde Q}_5=j q \left (
\begin{array}{ccc}
0 & 0 & j^2    \\
j & 0 & 0  \\
0 & 1 & 0    \\
\end{array}
\right)=j^2Q_5,
\end{eqnarray}

\begin{eqnarray}
{\tilde Q}_6= \left (
\begin{array}{ccc}
0 & 0 & j    \\
j^2 & 0 & 0  \\
0 & 1 & 0    \\
\end{array}
\right)=j^2Q_6
\nonumber\\
\end{eqnarray}

\begin{eqnarray}
Q_7= j \left (
\begin{array}{ccc}
j & 0 & 0    \\
0 & j^2 & 0  \\
0 & 0 & 1    \\
\end{array}
\right),
\end{eqnarray}

\begin{eqnarray}
Q_8=j^2 \left (
\begin{array}{ccc}
j^2 & 0   & 0    \\
0 & j & 0  \\
0 & 0   & 1    \\
\end{array}
\right),
\end{eqnarray}

\begin{eqnarray}
&&Q_7Q_1= q Q_2,\qquad Q_7Q_2= q Q_3,\qquad Q_7Q_3= q Q_1
\nonumber\\
&&Q_8Q_2= q^2 Q_1,\qquad Q_8Q_3= q^2 Q_2,\qquad Q_8Q_1= q^2 Q_3.
\nonumber\\
\end{eqnarray}

\begin{eqnarray}
&&Q_7Q_4= q Q_6,\qquad Q_7Q_6= q Q_5,\qquad Q_7Q_5= q Q_4
\nonumber\\
&&Q_8Q_4= q^2 Q_5,\qquad Q_8Q_5= q^2 Q_6,\qquad Q_8Q_6= q^2 Q_4.
\nonumber\\
\end{eqnarray}

\begin{eqnarray}
q_0= \left (
\begin{array}{ccc}
1 & 0 & 0    \\
0 & 1 & 0  \\
0 & 0 & 1    \\
\end{array}
\right)
\nonumber\\
\end{eqnarray}

\section{Ternary TU(3)-algebra}

We can consider the $3\times 3$  matrix realization of $q-$ algebra:

\begin{eqnarray}
&&q_1= \left (
\begin{array}{ccc}
0 & 1 & 0\\
0 & 0 & 1\\
1 & 0 & 0\\
\end{array}
\right),
\,
q_2=
\left (
\begin{array}{ccc}
0   & 1 & 0 \\
0   & 0 & j  \\
j^2 & 0 & 0\\
\end{array}
\right),
\,q_3=
\left (
\begin{array}{ccc}
0 & 1 & 0    \\
0 & 0 & j^2  \\
j & 0 & 0    \\
\end{array}
\right),
\nonumber\\
&&q_4=
\left (
\begin{array}{ccc}
0 & 0 & 1    \\
1 & 0 & 0  \\
0 & 1 & 0    \\
\end{array}
\right),
\,q_5=
\left (
\begin{array}{ccc}
0 & 0 & j    \\
1 & 0 & 0  \\
0 & j^2 & 0    \\
\end{array}
\right),
\,q_6=
\left (
\begin{array}{ccc}
0 & 0 & j^2    \\
1 & 0 & 0  \\
0 & j & 0    \\
\end{array}
\right),
\nonumber\\
&&q_7= j
\left (
\begin{array}{ccc}
1 & 0 & 0    \\
0 & j & 0  \\
0 & 0 & j^2    \\
\end{array}
\right),
\,q_8=j^2
\left (
\begin{array}{ccc}
1 & 0   & 0    \\
0 & j^2 & 0  \\
0 & 0   & j    \\
\end{array}
\right),
\,q_0=
\left (
\begin{array}{ccc}
1 & 0 & 0    \\
0 & 1 & 0  \\
0 & 0 & 1    \\
\end{array}
\right)
\nonumber\\
\end{eqnarray}

which satisfy to the ternary algebra:

\begin{eqnarray}
\{A,B,C\}_{S_3}=ABC+BCA+CAB-BAC-ACB-CBA.
\end{eqnarray}
Here $j=\exp(2{\bf i} \pi/3)$ and
$S_3$ is the permutation group of three elements.

On the next  table we can give the ternary commutation relations
for the matrices $q_k$:

\begin{equation}
\{q_k,q_l,q_m\}_{S_3}=f_{klm}^n q_n.
\end{equation}
One can check that each triple commutator $\{q_k,q_l,q_m\}$,
defined by triple numbers, $\{klm\}$ with $k,l,m=0,1,2,...,8$,
gives just one matrix $q_n$ with the corresponding coefficient
$f_{klm}^n$ giving in the table:

The $q_k$ elements satisfy to the ternary algebra:

\begin{eqnarray}
\{A,B,C\}_{S_3}=ABC+BCA+CAB-BAC-ACB-CBA.
\end{eqnarray}
Here $j=\exp(2{\bf i} \pi/3)$ and
$S_3$ is the permutation group of three elements.

On the next  table we can give the ternary commutation relations
for the matrices $q_k$:

\begin{equation}
\{q_k,q_l,q_m\}_{S_3}=f_{klm}^n q_n.
\end{equation}
One can check that each triple commutator $\{q_k,q_l,q_m\}$,
defined by triple numbers, $\{klm\}$ with $k,l,m=0,1,2,...,8$,
gives just one matrix $q_n$ with the corresponding coefficient
$f_{klm}^n$ giving in the table:

\begin{table}
\centering \caption{ \it  The ternary commutation relations}
\label{COM} \vspace{.05in}
\begin{tabular}{|c|c|c||c|c|c||c|c|c|}
\hline $   N     $&$  \{klm\} \rightarrow \{n\}  $&$  f_{klm}^n
$&$ N     $&$  \{klm\} \rightarrow  \{n\} $&$  f_{klm}^n $&$ N
$&$  \{klm\} \rightarrow \{n\}  $&$  f_{klm}^n $\\ \hline\hline $
1     $&$  \{123\} \rightarrow \{0\}  $&$  3(j^2-j) $&$ 2     $&$
\{124\} \rightarrow \{2\}  $&$  j(1-j) $&$ 3     $&$  \{125\}
\rightarrow \{1\}  $&$  2(j^2-j) $\\ \hline $   4     $&$  \{126\}
\rightarrow \{3\}  $&$  j(1-j) $&$ 5     $&$  \{127\} \rightarrow
\{5\}  $&$  2(1-j) $&$ 6     $&$  \{128\} \rightarrow \{4\}  $&$
2(j^2-1) $\\ \hline $   7     $&$  \{120\} \rightarrow \{6\}  $&$
(j^2-j) $&$ 8     $&$  \{134\} \rightarrow \{3\}  $&$  (j^2-j) $&$
9     $&$  \{135\} \rightarrow \{2\}  $&$  2(j-j^2) $\\ \hline $
10     $&$  \{136\} \rightarrow \{1\}  $&$  (j^2-j) $&$ 11     $&$
\{137\} \rightarrow \{4\}  $&$  2(j-1) $&$ 12     $&$  \{138\}
\rightarrow \{6\}  $&$  2(1-j^2) $\\ \hline $   13     $&$
\{130\} \rightarrow \{5\}  $&$  (j-j^2) $&$ 14     $&$  \{145\}
\rightarrow \{5\}  $&$ (j-j^2) $&$ 15     $&$  \{146\} \rightarrow
\{6\}  $&$  (j^2-j) $\\ \hline $   16     $&$  \{147\} \rightarrow
\{7\}  $&$  (j^2-j) $&$ 17     $&$  \{148\} \rightarrow \{8\}  $&$
(j-j^2) $&$ 18     $&$  \{140\} \rightarrow \tilde O  $&$    0 $\\
\hline $   19     $&$  \{156\} \rightarrow \{4\}  $&$  2j(j-1) $&$
20     $&$  \{157\} \rightarrow \{0\}  $&$  3(1-j) $&$ 21     $&$
\{158\} \rightarrow \{7\}  $&$  2(1-j) $\\ \hline $   22     $&$
\{150\} \rightarrow \{8\}  $&$  (1-j) $&$ 23     $&$  \{167\}
\rightarrow \{8\}  $&$  2(1-j^2) $&$ 24     $&$  \{168\}
\rightarrow \{0\}  $&$  3(1-j^2) $\\ \hline $   25     $&$
\{160\} \rightarrow \{7\}  $&$  (1-j^2) $&$ 26     $&$  \{178\}
\rightarrow \{1\}  $&$  (j-j^2) $&$ 27     $&$  \{170\}
\rightarrow \{2\}  $&$  (j-1) $\\ \hline $   28     $&$  \{180\}
\rightarrow \{3\}  $&$  (j^2-1) $&$ 29     $&$  \{234\}
\rightarrow \{1\}  $&$  2(j^2-j) $&$ 30     $&$  \{235\}
\rightarrow \{3\}  $&$  (j-j^2) $\\ \hline $   31     $&$  \{236\}
\rightarrow \{2\}  $&$  (j-j^2) $&$ 32     $&$  \{237\}
\rightarrow \{6\}  $&$  2(1-j) $&$ 33     $&$  \{238\} \rightarrow
\{5\}  $&$  2(j^2-1) $\\ \hline $   34     $&$  \{230\}
\rightarrow \{4\}  $&$  (j^2-j) $&$ 35     $&$  \{245\}
\rightarrow \{4\}  $&$  (j-j^2) $&$ 36     $&$  \{246\}
\rightarrow \{5\}  $&$  2(j-j^2) $\\ \hline $   37     $&$
\{247\} \rightarrow \{8\}  $&$  2(1-j^2) $&$ 38     $&$  \{248\}
\rightarrow \{0\}  $&$  3(1-j^2) $&$ 39     $&$  \{240\}
\rightarrow \{7\}  $&$  (1-j^2) $\\ \hline $   40     $&$  \{256\}
\rightarrow \{6\}  $&$  (j-j^2) $&$ 41     $&$  \{257\}
\rightarrow \{7\}  $&$  (j^2-j) $&$ 42     $&$  \{258\}
\rightarrow \{8\}  $&$  (j-j^2) $\\ \hline $   43     $&$  \{250\}
\rightarrow \tilde O  $&$  0 $&$ 44     $&$  \{267\} \rightarrow
\{0\}  $&$  3(1-j) $&$ 45     $&$  \{268\} \rightarrow \{7\}  $&$
2(1-j) $\\ \hline $   46     $&$  \{260\} \rightarrow \{8\}  $&$
(1-j) $&$ 47     $&$  \{278\} \rightarrow \{2\}  $&$  (j-j^2) $&$
48     $&$  \{270\} \rightarrow \{3\}  $&$  (j-1) $\\ \hline $
49     $&$  \{280\} \rightarrow \{1\}  $&$  (j^2-1) $&$ 50     $&$
\{345\} \rightarrow \{6\}  $&$  2(j^2-j) $&$ 51     $&$  \{346\}
\rightarrow \{4\}  $&$  (j^2-j) $\\ \hline $   52     $&$  \{347\}
\rightarrow \{0\}  $&$  3(1-j) $&$ 53     $&$  \{348\} \rightarrow
\{7\}  $&$  2(1-j) $&$ 54     $&$  \{340\} \rightarrow \{8\}  $&$
(1-j) $\\ \hline $   55     $&$  \{356\} \rightarrow \{5\}  $&$
j-j^2 $&$ 56     $&$  \{357\} \rightarrow \{8\}  $&$  2(1-j^2) $&$
57     $&$  \{358\} \rightarrow \{0\}  $&$  3(1-j^2) $\\ \hline $
58     $&$  \{350\} \rightarrow \{7 \} $&$  (1-j^2) $&$ 59     $&$
\{367\} \rightarrow \{7\}  $&$  (j^2-j) $&$ 60     $&$  \{368\}
\rightarrow \{8\}  $&$  (j-j^2) $\\ \hline $   61     $&$  \{360\}
\rightarrow \tilde O $&$   0 $&$ 62     $&$  \{378\} \rightarrow
\{3\}  $&$  (j-j^2) $&$ 63     $&$  \{370\} \rightarrow \{1\}  $&$
(j-1) $\\ \hline $   64     $&$  \{380\} \rightarrow \{2\} $&$
(j^2-1) $&$ 65     $&$  \{456\} \rightarrow \{0\}  $&$  3(j^2-j)
$&$ 66     $&$  \{457\} \rightarrow \{1\}  $&$  2(1-j) $\\ \hline
$   67     $&$  \{458\} \rightarrow \{2\}  $&$   2(j^2-1) $&$ 68
$&$  \{450\} \rightarrow \{3\}  $&$  (j^2-j) $&$ 69     $&$
\{467\} \rightarrow \{1\}  $&$  2(j-1) $\\ \hline $   70     $&$
\{468\} \rightarrow \{1\} $&$   2(1-j^2) $&$ 71     $&$  \{460\}
\rightarrow \{2\}  $&$  (j-j^2) $&$ 72     $&$  \{478\}
\rightarrow \{4\}  $&$  (j^2-j) $\\ \hline $   73     $&$  \{470\}
\rightarrow \{6\}  $&$  (1-j) $&$ 74     $&$  \{480\} \rightarrow
\{5\}  $&$  (1-j^2) $&$ 75     $&$  \{567\} \rightarrow \{2\}  $&$
2(1-j) $\\ \hline $   76     $&$  \{568\} \rightarrow \{3\}  $&$
2(j^2-j) $&$ 77     $&$  \{560\} \rightarrow \{1\}  $&$  (j^2-j)
$&$ 78     $&$  \{578\} \rightarrow \{5\}  $&$  (j^2-j) $\\ \hline
$   79     $&$  \{570\} \rightarrow \{4\}  $&$  (1-j) $&$ 80
$&$  \{580\} \rightarrow \{6\}  $&$  (1-j^2) $&$ 81     $&$
\{678\} \rightarrow \{6\}  $&$  (j^2-j) $\\ \hline $   82     $&$
\{670\} \rightarrow \{5\}  $&$   (1-j) $&$ 83     $&$  \{680\}
\rightarrow \{4\}  $&$  (1-j^2) $&$ 84     $&$  \{780\}
\rightarrow \tilde O $&$   0 $\\ \hline
\end{tabular}
\end{table}

One can find $C_9^2=84$ ternary commutation relations. But there one  can see
that there are also $C_8^2=28 $commutation relations which
correspond to the $su(3)$ algebra! Therefore, it is naturally
to represent the q-numbers as  ternary generalization of quaternions.
If one can take from $S_3$ commutation relations $C=q_0$ the
commutation relations naturally are going to $S_2$ Lie commutation
relations:

\begin{equation}
\{q_a,q_b, q_0\}_{S_3}=q_aq_bq_0+q_bq_0q_a+q_0q_aq_b-q_bq_aq_0
-q_aq_0q_b-q_0q_bq_a=q_aq_b - q_bq_a,
\end{equation}
where $a \neq b \neq 0$.
On the table such 28- cases one can see $\{kl0\}$.

We can consider the $3\times 3$  matrix realization of $q-$ algebra:

\begin{eqnarray}
&&q_1=q \left (
\begin{array}{ccc}
0 & 1 & 0\\
0 & 0 & 1\\
1 & 0 & 0\\
\end{array}
\right), \, q_2=q^2 \left (
\begin{array}{ccc}
0   & 1 & 0 \\
0   & 0 & j  \\
j^2 & 0 & 0\\
\end{array}
\right),
\,q_3=
\left (
\begin{array}{ccc}
0 & 1 & 0    \\
0 & 0 & j^2  \\
j & 0 & 0    \\
\end{array}
\right),
\nonumber\\
&&q_4=q^2 \left (
\begin{array}{ccc}
0 & 0 & 1    \\
1 & 0 & 0  \\
0 & 1 & 0    \\
\end{array}
\right), \,q_5=q \left (
\begin{array}{ccc}
0 & 0 & j    \\
1 & 0 & 0  \\
0 & j^2 & 0    \\
\end{array}
\right),
\,q_6=
\left (
\begin{array}{ccc}
0 & 0 & j^2    \\
1 & 0 & 0  \\
0 & j & 0    \\
\end{array}
\right),
\nonumber\\
&&q_7=q^2 \left (
\begin{array}{ccc}
1 & 0 & 0    \\
0 & j & 0  \\
0 & 0 & j^2    \\
\end{array}
\right), \,q_8=q \left (
\begin{array}{ccc}
1 & 0   & 0    \\
0 & j^2 & 0  \\
0 & 0   & j    \\
\end{array}
\right),
\,q_0=
\left (
\begin{array}{ccc}
1 & 0 & 0    \\
0 & 1 & 0  \\
0 & 0 & 1    \\
\end{array}
\right)
\nonumber\\
\end{eqnarray}

which satisfy to the ternary algebra:

\begin{eqnarray}
\{A,B,C\}_{S_3}=ABC+BCA+CAB-BAC-ACB-CBA.
\end{eqnarray}
Here $j=\exp(2{\bf i} \pi/3)$ and
$S_3$ is the permutation group of three elements.

On the next  table we can give the ternary commutation relations
for the matrices $q_k$:

\begin{equation}
\{q_k,q_l,q_m\}_{S_3}=f_{klm}^n q_n.
\end{equation}
One can check that each triple commutator $\{q_k,q_l,q_m\}$,
defined by triple numbers, $\{klm\}$ with $k,l,m=0,1,2,...,8$,
gives just one matrix $q_n$ with the corresponding coefficient
$f_{klm}^n$ giving in the table:

\section{The geometrical representations  of ternary   "quaternions"}

Let us define the following product:

\begin{eqnarray}
\hat Q= \sum_{a=0}^{a=8} \{x_a q_a\}
 &=&\left (
\begin{array}{ccc}
x_0+j x_7+j^2x_8 & x_1+x_2+x_3     & x_4+jx_5+j^2x_6\\
x_4+x_5+x_6      & x_0+j^2x_7+jx_8 & x_1+jx_2+j^2x_3\\
x_1+j^2x_2+jx_3  & x_4+j^2x_5+jx_6 & x_0 +x_7+x_8   \\
\end{array}
\right) \nonumber\\
&=&\left (
\begin{array}{ccc}
{\tilde z}_0 & z_1     & {\tilde z}_2\\
z_2      & {\tilde {\tilde z}}_0 & {\tilde z}_1\\
 {\tilde {\tilde z}}_1  & {\tilde {\tilde z}}_2 & z_0   \\
\end{array}
\right) .
\end{eqnarray}

\begin{eqnarray}
&&(z_0z_1z_2+{\tilde z}_0{\tilde z}_1{\tilde z}_2+ {\tilde {\tilde
z}}_0  {\tilde {\tilde z}}_1  {\tilde {\tilde z}}_2  ) \nonumber\\
&=& [(x_0 +x_7+x_8)(x_1+ x_2+x_3 )(x_4+x_5+x_6)]\nonumber\\
&+&[(x_0+jx_7+j^2x_8)(x_1+jx_2+j^2x_3)(x_4+j^2x_5+jx_6)]       \nonumber\\
&+&[(x_0+j^2x_7+jx_8)(x_1+j^2x_2+jx_3)(x_4+jx_5+j^2x_6)]       \nonumber\\
&=& [(x_0 +x_7+x_8) \nonumber\\
&&\cdot (x_1x_4+x_1x_5+x_1x_6+x_2x_4+x_2x_5+x_2x_6+x_3x_4+x_3x_5+x_3x_6]\nonumber\\
&+&[(x_0 +jx_7+j^2x_8) \nonumber\\
&&\cdot (x_1x_4+j^2x_1x_5+jx_1x_6+jx_2x_4+x_2x_5+j^2x_2x_6+j^2x_3x_4+jx_3x_5+x_3x_6]\nonumber\\
&+&[(x_0 +j^2x_7+jx_8) \nonumber\\
&&\cdot (x_1x_4+jx_1x_5+j^2x_1x_6+j^2x_2x_4+x_2x_5+jx_2x_6+jx_3x_4+j^2x_3x_5+x_3x_6]\nonumber\\
&=&\{3 x_0 [x_1x_4+x_2x_5+x_3x_6]\}\nonumber\\
&+&\{3x_7[x_1x_5+x_2x_6+x_3x_4]\}   \nonumber\\
&+&\{3x_8 [x_1x_6+x_2x_4+x_3x_5]\} \nonumber\
\end{eqnarray}

Then we can define the norm of the ternary quaternion through the determinant
\begin{eqnarray}
Det \hat Q&=&[(x_0+j x_7+j^2x_8 )(x_0+j^2x_7+jx_8)(x_0 +x_7+x_8)] \nonumber\\
&+&[(x_1+x_2+x_3) (x_1+jx_2+j^2x_3)(x_1+j^2x_2+jx_3)]        \nonumber\\
&+&[(x_4+x_5+x_6)  (x_4+jx_5+j^2x_6)(x_4+j^2x_5+jx_6)]       \nonumber\\
&-&\{(x_0+j^2x_7+jx_8)(x_1+j^2x_2+jx_3 )(x_4+jx_5+j^2x_6)\nonumber\\
&-&(x_0+j x_7+j^2x_8) (x_1+jx_2+j^2x_3)(x_4+j^2x_5+jx_6 )\nonumber\\
&-&(x_0 +x_7+x_8  )(x_1+x_2+x_3) (x_4+x_5+x_6 )\} \nonumber\\
&=&|z_0|^3+|z_1|^3+|z_2|^3-(z_0z_1z_2+{\tilde z}_0{\tilde
z}_1{\tilde z}_2+ {\tilde {\tilde z}}_0  {\tilde {\tilde z}}_1  {\tilde {\tilde z}}_2  )\nonumber\\
\end{eqnarray}

\begin{eqnarray}
\begin{array}{ccc}
z_0=x_0+x_7q+x_8q^2 &{\tilde z}_0=x_0+jx_7q+j^2x_8q^2& {\tilde{\tilde z}}_0=x_0+j^2x_7q+jx_8q^2\\
z_1=x_1+x_2q+x_3q^2 &{\tilde z}_1=x_1+jx_2q+j^2x_3q^2& {\tilde{\tilde z}}_1=x_1+j^2x_2q+jx_3q^2\\
z_2=x_4+x_5q+x_6q^2 &{\tilde z}_2=x_4+jx_5q+j^2x_6q^2& {\tilde{\tilde z}}_2=x_4+j^2x_5q+jx_6q^2\\
\end{array}
\end{eqnarray}
or
\begin{eqnarray}
Det \hat Q&=&[x_0^3+x_7^3+x_8^3-3x_0x_7x_8] \nonumber\\
&+&[ x_1^3+x_2^3+x_3^3-3x_1x_2x_3]      \nonumber\\
&+&[(x_4^3+x_5^3+x_6^3-3 x_4x_5x_6)]       \nonumber\\
&-&\{ (x_0+j^2x_7+jx_8)\nonumber\\
&\cdot&[x_1x_4+jx_1x_5+j^2x_1x_6+j^2x_2x_4+x_2x_5+jx_2x_6
+jx_3x_4+j^2x_3x_5+x_3x_6 ]\}\nonumber\\
&-&\{(x_0+x_7+x_8) \nonumber\\
&\cdot&[x_1x_4+j^2x_1x_5+jx_1x_6+jx_2x_4+x_2x_5+j^2x_2x_6
+j^2x_3x_4+jx_3x_5+x_3x_6]\}\nonumber\\
&-&\{(x_0+j x_7+j^2x_8) \nonumber\\
  &\cdot &[ x_1x_4+x_1x_5+x_1x_6+x_2x_4+x_2x_5+x_2x_6
+x_3x_4+x_3x_5+x_3x_6]\} \nonumber\\
\end{eqnarray}

\begin{eqnarray}
\begin{array}{ccc}
z_0=x_0+x_7q+x_8q^2 &{\tilde z}_0=x_0+jx_7q+j^2x_8q^2& {\tilde{\tilde z}}_0=x_0+j^2x_7q+jx_8q^2\\
z_1=x_1+x_2q+x_3q^2 &{\tilde z}_1=x_1+jx_2q+j^2x_3q^2& {\tilde{\tilde z}}_1=x_1+j^2x_2q+jx_3q^2\\
z_2=x_4+x_5q+x_6q^2 &{\tilde z}_2=x_4+jx_5q+j^2x_6q^2& {\tilde{\tilde z}}_2=x_4+j^2x_5q+jx_6q^2\\
\end{array}
\end{eqnarray}
or(1)

\begin{eqnarray}
Det \hat Q&=&[x_0^3+x_7^3+x_8^3-3x_0x_7x_8] \nonumber\\
&+&[ x_1^3+x_2^3+x_3^3-3x_1x_2x_3]      \nonumber\\
&+&[(x_4^3+x_5^3+x_6^3-3 x_4x_5x_6)]       \nonumber\\
&-&\{3 x_0 [x_1x_4+x_2x_5+x_3x_6]\}\nonumber\\
&-&\{3x_7 [x_1x_5+x_2x_6+x_3x_4]\}\nonumber\\
&-&\{3x_8 [ x_1x_6+x_2x_4+x_3x_5]\} \nonumber\\
\end{eqnarray}
or (2)

\begin{eqnarray}
Det \hat Q&=&[x_0^3+x_1^3+x_4^3-3x_0x_1x_4] \nonumber\\
&+&[ x_7^3+x_2^3+x_6^3-3x_7x_2x_6]      \nonumber\\
&+&[(x_8^3+x_5^3+x_3^3-3 x_8x_5x_3)]       \nonumber\\
&-&\{3 x_0 [x_7x_8+x_2x_5+x_3x_6]\}\nonumber\\
&-&\{3x_1 [x_2x_3+x_5x_7+x_6x_8]\}\nonumber\\
&-&\{3x_4 [ x_5x_6+x_3x_7+x_2x_8]\} \nonumber\\
\end{eqnarray}

\begin{eqnarray}
\begin{array}{ccc}
z_0=x_0+x_1q+x_4q^2 &{\tilde z}_0=x_0+jx_1q+j^2x_4^2& {\tilde{\tilde z}}_0=x_0+j^2x_1q+jx_4q^2\\
z_1=x_7+x_2q+x_6q^2 &{\tilde z}_1=x_7+jx_2q+j^2x_6q^2& {\tilde{\tilde z}}_1=x_7+j^2x_2q+jx_6q^2\\
z_2=x_8+x_5q+x_3q^2 &{\tilde z}_2=x_8+jx_5q+j^2x_3q^2& {\tilde{\tilde z}}_2=x_8+j^2x_5q+jx_3q^2\\
\end{array}
\end{eqnarray}

or (3)
\begin{eqnarray}
Det \hat Q&=&[x_0^3+x_2^3+x_5^3-3x_0x_2x_5] \nonumber\\
&+&[ x_7^3+x_3^3+x_4^3-3x_7x_3x_4]      \nonumber\\
&+&[(x_8^3+x_1^3+x_6^3-3 x_8x_1x_6)]       \nonumber\\
&-&\{3 x_0 [x_7x_8+x_1x_4+x_3x_6]\}\nonumber\\
&-&\{3x_2 [x_1x_3+x_6x_7+x_4x_8]\}\nonumber\\
&-&\{3x_5 [ x_3x_4+x_1x_7+x_3x_8]\} \nonumber\\
\end{eqnarray}

\begin{eqnarray}
\begin{array}{ccc}
z_0=x_0+x_2q+x_5q^2 &{\tilde z}_0=x_0+jx_2q+j^2x_5q^2& {\tilde{\tilde z}}_0=x_0+j^2x_2q+jx_5q^2\\
z_1=x_7+x_3q+x_4q^2 &{\tilde z}_1=x_7+jx_3q+j^2x_4q^2& {\tilde{\tilde z}}_1=x_7+j^2x_3q+jx_4q^2\\
z_2=x_8+x_1q+x_6q^2 &{\tilde z}_2=x_8+jx_1q+j^2x_6q^2& {\tilde{\tilde z}}_2=x_8+j^2x_1q+jx_6q^2\\
\end{array}
\end{eqnarray}

or (4)
\begin{eqnarray}
Det \hat Q&=&[x_0^3+x_3^3+x_6^3-3x_0x_3x_6] \nonumber\\
&+&[ x_7^3+x_1^3+x_5^3-3x_1x_5x_7]      \nonumber\\
&+&[(x_8^3+x_2^3+x_4^3-3 x_2x_4x_8)]       \nonumber\\
&-&\{3 x_0 [x_7x_8+x_1x_4+x_2x_5]\}\nonumber\\
&-&\{3x_3 [x_1x_2+x_4x_7+x_5x_8]\}\nonumber\\
&-&\{3x_6 [ x_4x_5+x_2x_7+x_1x_8]\} \nonumber\\
\end{eqnarray}

\begin{eqnarray}
\begin{array}{ccc}
z_0=x_0+x_3q+x_6q^2 &{\tilde z}_0=x_0+jx_3q+j^2x_6q^2& {\tilde{\tilde z}}_0=x_0+j^2x_3q+jx_6q^2\\
z_1=x_7+x_1q+x_5q^2 &{\tilde z}_1=x_7+jx_1q+j^2x_5q^2& {\tilde{\tilde z}}_1=x_7+j^2x_1q+jx_5q^2\\
z_2=x_8+x_2q+x_4q^2 &{\tilde z}_2=x_8+jx_2q+j^2x_4q^2& {\tilde{\tilde z}}_2=x_8+j^2x_2q+jx_4q^2\\
\end{array}
\end{eqnarray}

\begin{eqnarray}
\begin{array}{ccc}
0 & 8  & 7 \\
1 & 2  & 3 \\
4 & 6  & 5 \\
\end{array}
\qquad
\begin{array}{ccc}
0 & 8  & 7 \\
2 & 3  & 1 \\
5 & 4  & 6 \\
\end{array}
\qquad
\begin{array}{ccc}
0 & 8  & 7 \\
3 & 1  & 2 \\
6 & 5  & 4 \\
\end{array}
\qquad
\end{eqnarray}

\begin{eqnarray}
\begin{array}{ccc}
0 & 1  & 4 \\
2 & 7  & 6 \\
5 & 3  & 8 \\
\end{array}
\qquad
\begin{array}{ccc}
0 & 1  & 4 \\
6 & 2  & 7 \\
3 & 8  & 5 \\
\end{array}
\qquad
\begin{array}{ccc}
0 & 1  & 4 \\
7 & 6  & 2 \\
8 & 5  & 3 \\
\end{array}
\qquad
\end{eqnarray}

\begin{eqnarray}
\begin{array}{ccc}
0 & 2  & 5 \\
3 & 7  & 4 \\
6 & 1  & 8 \\
\end{array}
\qquad
\begin{array}{ccc}
0 & 2  & 5 \\
4 & 3  & 7 \\
1 & 8  & 6 \\
\end{array}
\qquad
\begin{array}{ccc}
0 & 2  & 5 \\
7 & 4  & 3 \\
8 & 6  & 1 \\
\end{array}
\qquad
\end{eqnarray}

\begin{eqnarray}
\begin{array}{ccc}
0 & 3  & 6 \\
1 & 7  & 5 \\
4 & 2  & 8 \\
\end{array}
\qquad
\begin{array}{ccc}
0 & 3  & 6 \\
5 & 1  & 7 \\
2 & 8  & 4 \\
\end{array}
\qquad
\begin{array}{ccc}
0 & 3  & 6 \\
7 & 5  & 1 \\
8 & 4  & 2 \\
\end{array}
\qquad
\end{eqnarray}

 One can see that this norm is a real number and if we define
this norm to unit $Det \hat Q=1$, it will define a cubic surface
in $D=9$.

\section{Real ternary Tu3-algebra and root system}

 Let us give the link the nonions
with the canonical $SU(3) $ matrices:

\begin{eqnarray}
\lambda_1=\left(
\begin{array}{ccc}
0 & 1 & 0 \\
1 & 0 & 0 \\
0 & 0 & 0 \\
\end{array}
\right)
=
\frac{1}{3}(q_1+q_2+q_3+q_4+jq_5+q_6)
\end{eqnarray}
\begin{eqnarray}
\lambda_2=\left(
\begin{array}{ccc}
0 &-i & 0 \\
i & 0 & 0 \\
0 & 0 & 0 \\
\end{array}
\right)
=
\frac{i}{3}(-q_1-q_2-q_3+q_4+jq_5+j^2q_6)
\end{eqnarray}

\begin{eqnarray}
\lambda_4=\left(
\begin{array}{ccc}
0 & 0 & 0 \\
0 & 0 & 1 \\
0 & 1 & 0 \\
\end{array}
\right)
=
\frac{1}{3}(q_1+j^2q_2+jq_3+q_4+jq_5+j^2q_6)
\end{eqnarray}

\begin{eqnarray}
\lambda_5=\left(
\begin{array}{ccc}
0 & 0 & 0 \\
0 & 0 & -i \\
0 & i & 0 \\
\end{array}
\right)
=
\frac{i}{3}(-q_1-j^2q_2-jq_3+q_4+jq_5+j^2q_6)
\end{eqnarray}

\begin{eqnarray}
\lambda_6=\left(
\begin{array}{ccc}
0 & 0 & 1 \\
0 & 0 & 0 \\
1 & 0 & 0 \\
\end{array}
\right)
=
\frac{i}{3}(q_1+j q_2+j^2 q_3+q_4+j^2q_5+jq_6)
\end{eqnarray}
\begin{eqnarray}
\lambda_7=\left(
\begin{array}{ccc}
0 & 0 & -i \\
0 & 0 & 0 \\
i & 0 & 0 \\
\end{array}
\right)
=
\frac{i}{3}(q_1+jq_2+j^2q_3-q_4-j^2q_5-jq_6)
\end{eqnarray}

\begin{eqnarray}
\lambda_3=\left(
\begin{array}{ccc}
1 & 0 & 0 \\
0 & -1 & 0 \\
0 & 0 & 0 \\
\end{array}
\right)
=
\frac{1}{(1-j)}(q_7-jq_8)
\end{eqnarray}

\begin{eqnarray}
\lambda_8=\frac{1}{\sqrt{3}}\left(
\begin{array}{ccc}
1 & 0 & 0 \\
0 & 1 & 0 \\
0 & 0 & -2 \\
\end{array}
\right)
=
\frac{-1}{\sqrt{3}}(jq_7+j^2q_8)
\end{eqnarray}

Here the matrices $\lambda_i/2=g_i$ satisfy to ordinary $SU(3)$ algebra:

\begin{equation}
[g_i,g_j]_{Z_2}=if_{ijk}g_k.
\end{equation}
where $f_{ijk}$ are completely antisymmetric and have the following values:
\begin{equation}
f_{123}=1,
f_{147}=f_{165}=f_{246}=f_{257}=f_{345}=f_{376}=\frac{1}{2},
f_{458}=f_{678}=\frac{\sqrt{3}}{2}.
\end{equation}

Now we introduce the plus-step operators:

\begin{scriptsize}
\begin{eqnarray}
Q_1=Q_I^+=\left(
\begin{array}{ccc}
0 & 1 & 0 \\
0 & 0 & 0 \\
0 & 0 & 0 \\
\end{array}
\right)
\qquad
Q_2=Q_{II}^+=\left(
\begin{array}{ccc}
0 & 0 & 0 \\
0 & 0 & 1 \\
0 & 0 & 0 \\
\end{array}
\right)
\qquad
Q_3=Q_{III}^+=\left(
\begin{array}{ccc}
0 & 0 & 0 \\
0 & 0 & 0 \\
1 & 0 & 0 \\
\end{array}
\right)
\end{eqnarray}
\end{scriptsize}

and on the minus-step operators:
\begin{scriptsize}
\begin{eqnarray}
Q_4=Q_{I}^-=\left(
\begin{array}{ccc}
0 & 0 & 0 \\
1 & 0 & 0 \\
0 & 0 & 0 \\
\end{array}
\right)
\qquad
Q_5 =Q_{II}^-=\left(
\begin{array}{ccc}
0 & 0 & 0 \\
0 & 0 & 0 \\
0 & 1 & 0 \\
\end{array}
\right)
\qquad
Q_6=Q_{III}^-=\left(
\begin{array}{ccc}
0 & 0 & 1 \\
0 & 0 & 0 \\
0 & 0 & 0 \\
\end{array}
\right)
\end{eqnarray}
\end{scriptsize}

We choose the following 3-diagonal
operators :

\begin{scriptsize}
\begin{eqnarray}
Q_7=Q_{I}^0=\left(
\begin{array}{ccc}
\frac{1}{\sqrt{6}} &       0            &       0                   \\
0                  & \frac{1}{\sqrt{6}} &       0                   \\
0                  &        0           & -\sqrt{\frac{2}{3}} \\
\end{array}
\right)
\qquad
Q_8=Q_{II}^0=\left(
\begin{array}{ccc}
\frac{1}{\sqrt{2}} & 0                   & 0 \\
0                  & -\frac{1}{\sqrt{2}} & 0 \\
0                  & 0                   & 0 \\
\end{array}
\right)
\qquad
Q_0=Q_{III}^0=\left(
\begin{array}{ccc}
\frac{1}{\sqrt{3}}& 0                  & 0                 \\
0                 &  \frac{1}{\sqrt{3}}& 0                 \\
0                 & 0                  & \frac{1}{\sqrt{3}}\\
\end{array}
\right),
\end{eqnarray}
\end{scriptsize}
which produce the ternary Cartan subalgebra:
\begin{eqnarray}
\{Q_0,Q_7,Q_8 \}=0.
\end{eqnarray}

The $Q_k$ operators with $k=0,1,2,...,8$  satisfy to the following ternary $S_3$
commutation relations: \\
\vspace{0.1in}
\noindent\(
\begin{tiny}
\begin{array}{|c|c|c||c|c|c||c|c|c|} \hline
   N & \{klm\} \rightarrow  \{n\} & f_{klm}^n
 & N & \{klm\} \rightarrow  \{n\} & f_{klm}^n
 & N & \{klm\} \rightarrow  \{n\} & f_{klm}^n
 \\ \hline  \hline
   1 & \{0,1,2\}\to         \{6\} & \frac{1}{\sqrt{3}}
 &29 & \{1,2,3\}\to         \{0\} & \sqrt{3}
 &57 & \{2,4,7\}\to    \emptyset  & 0
 \\ \hline
 2 & \{0,1,3\}\to \{5\} & -\frac{1}{\sqrt{3}} & 30 & \{1,2,4\}\to \{2\} & 1 & 58 & \{2,4,8\}\to \emptyset  & 0 \\ \hline
 3 & \{0,1,4\}\to \{0,7,8\} & \left\{0,0,\sqrt{\frac{2}{3}} \right \}& 31 & \{1,2,5\}\to \{1\} & 1 & 59 & \{2,5,6\}\to \{6\} & -1 \\ \hline
 4 & \{0,1,5\}\to \emptyset  & 0 & 32 & \{1,2,6\}\to \emptyset  & 0 & 60 & \{2,5,7\}\to \{0,7,8\} & \left\{\frac{3}{\sqrt{2}},-1,-\frac{2}{\sqrt{3}}\right\}
\\ \hline
 5 & \{0,1,6\}\to \emptyset  & 0 & 33 & \{1,2,7\}\to \{6\} & -\sqrt{\frac{2}{3}} & 61 & \{2,5,8\}\to \{0,7,8\} & \left\{-\sqrt{\frac{3}{2}},0,1\right\}
\\ \hline
 6 & \{0,1,7\}\to \emptyset  & 0 & 34 & \{1,2,8\}\to \{6\} & \sqrt{2} & 62 & \{2,6,7\}\to \emptyset  & 0 \\ \hline
 7 & \{0,1,8\}\to \{1\} & -\sqrt{\frac{2}{3}} & 35 & \{1,3,4\}\to \{3\} & -1 & 63 & \{2,6,8\}\to \emptyset  & 0 \\ \hline
 8 & \{0,2,3\}\to \{4\} & \frac{1}{\sqrt{3}} & 36 & \{1,3,5\}\to \emptyset  & 0 & 64 & \{2,7,8\}\to \{2\} & \frac{1}{\sqrt{3}} \\ \hline
 9 & \{0,2,4\}\to \emptyset  & 0 & 37 & \{1,3,6\}\to \{1\} & -1 & 65 & \{3,4,5\}\to \emptyset  & 0 \\ \hline
 10 & \{0,2,5\}\to \{0,7,8\} & \left\{0,\frac{1}{\sqrt{2}},-\frac{1}{\sqrt{6}}\right\} & 38 & \{1,3,7\}\to \{5\} & \sqrt{\frac{2}{3}} & 66 & \{3,4,6\}\to
\{4\} & 1 \\ \hline
 11 & \{0,2,6\}\to \emptyset  & 0 & 39 & \{1,3,8\}\to \{5\} & \sqrt{2} & 67 & \{3,4,7\}\to \emptyset  & 0 \\ \hline
 12 & \{0,2,7\}\to \{2\} & -\frac{1}{\sqrt{2}} & 40 & \{1,4,5\}\to \{5\} & -1 & 68 & \{3,4,8\}\to \emptyset  & 0 \\ \hline
 13 & \{0,2,8\}\to \{2\} & \frac{1}{\sqrt{6}} & 41 & \{1,4,6\}\to \{6\} & 1 & 69 & \{3,5,6\}\to \{5\} & -1 \\ \hline
 14 & \{0,3,4\}\to \emptyset  & 0 & 42 & \{1,4,7\}\to \{0,7,8\} & \left\{0,0,\frac{1}{\sqrt{3}} \right\}& 70 & \{3,5,7\}\to \emptyset  & 0 \\ \hline
 15 & \{0,3,5\}\to \emptyset  & 0 & 43 & \{1,4,8\}\to \{0,7,8\} & \left\{\sqrt{6},\sqrt{3},0\right\} & 71 & \{3,5,8\}\to \emptyset  & 0 \\ \hline
 16 & \{0,3,6\}\to \{0,7,8\} & \left\{0,-\frac{1}{\sqrt{2}},-\frac{1}{\sqrt{6}}\right\} & 44 & \{1,5,6\}\to \emptyset  & 0 & 72 & \{3,6,7\}\to \{0,7,8\}
& \left\{-\frac{3}{\sqrt{2}},1,-\frac{2}{\sqrt{3}}\right\} \\ \hline
 17 & \{0,3,7\}\to \{3\} & \frac{1}{\sqrt{2}} & 45 & \{1,5,7\}\to \emptyset  & 0 & 73 & \{3,6,8\}\to \{0,7,8\} & \left\{-\sqrt{\frac{3}{2}},0,-1\right\}
\\ \hline
 18 & \{0,3,8\}\to \{3\} & \frac{1}{\sqrt{6}} & 46 & \{1,5,8\}\to \emptyset  & 0 & 74 & \{3,7,8\}\to \{3\} & \frac{1}{\sqrt{3}} \\ \hline
 19 & \{0,4,5\}\to \{3\} & -\frac{1}{\sqrt{3}} & 47 & \{1,6,7\}\to \emptyset  & 0 & 75 & \{4,5,6\}\to \{0\} & -\sqrt{3} \\ \hline
 20 & \{0,4,6\}\to \{2\} & \frac{1}{\sqrt{3}} & 48 & \{1,6,8\}\to \emptyset  & 0 & 76 & \{4,5,7\}\to \{3\} & \sqrt{\frac{2}{3}} \\ \hline
 21 & \{0,4,7\}\to \emptyset  & 0 & 49 & \{1,7,8\}\to \{1\} & \frac{1}{\sqrt{3}} & 77 & \{4,5,8\}\to \{3\} & -\sqrt{2} \\ \hline
 22 & \{0,4,8\}\to \{4\} & \sqrt{\frac{2}{3}} & 50 & \{2,3,4\}\to \emptyset  & 0 & 78 & \{4,6,7\}\to \{2\} & -\sqrt{\frac{2}{3}} \\ \hline
 23 & \{0,5,6\}\to \{1\} & -\frac{1}{\sqrt{3}} & 51 & \{2,3,5\}\to \{3\} & 1 & 79 & \{4,6,8\}\to \{2\} & -\sqrt{2} \\ \hline
 24 & \{0,5,7\}\to \{5\} & \frac{1}{\sqrt{2}} & 52 & \{2,3,6\}\to \{2\} & 1 & 80 & \{4,7,8\}\to \{4\} & -\frac{1}{\sqrt{3}} \\ \hline
 25 & \{0,5,8\}\to \{5\} & -\frac{1}{\sqrt{6}} & 53 & \{2,3,7\}\to \{4\} & 2 \sqrt{\frac{2}{3}} & 81 & \{5,6,7\}\to \{1\} & -2 \sqrt{\frac{2}{3}}
\\ \hline
 26 & \{0,6,7\}\to \{6\} & -\frac{1}{\sqrt{2}} & 54 & \{2,3,8\}\to \emptyset  & 0 & 82 & \{5,6,8\}\to \emptyset  & 0 \\ \hline
 27 & \{0,6,8\}\to \{6\} & -\frac{1}{\sqrt{6}} & 55 & \{2,4,5\}\to \{4\} & -1 & 83 & \{5,7,8\}\to \{5\} & -\frac{1}{\sqrt{3}} \\ \hline
 28 & \{0,7,8\}\to \emptyset  & 0 & 56 & \{2,4,6\}\to \emptyset  & 0 & 84 & \{6,7,8\}\to \{6\} & -\frac{1}{\sqrt{3}} \\ \hline
\end{array}
\end{tiny}
\)

We have got 84 commutations relations. One commutation relation, $\{Q_0,Q_7,Q_8\}_{S_3}=0$,
provide the Cartan subalgebra.
Let separate the rest 83 commutation relations on the 5 groups$(18+18+27+18+2)$.
The first group contains itself  the following 18 commutation relations in one group ( see Table):

\begin{scriptsize}
\begin{table}
\caption{ \it  I-The  root system of the step operators}
\label{COMWeyl} \vspace{.05in}
\begin{tabular}{||c|c||c|c||c|c|c|} \hline
$\{klm\} \rightarrow \{n\} $&$ f_{klm}^n $&$ \{klm\} \rightarrow  \{n\} $&$f_{klm}^n $&$ \{klm\} \rightarrow \{n\}$&$f_{klm}^n  $&$ \vec \alpha_i   $ \\ \hline \hline
$           \{1,7,8\}\to \{1\}  $&$\frac{1}{ \sqrt{3}}
$&$         \{2,7,8\}\to \{2\}  $&$\frac{1}{ \sqrt{3}}
$&$         \{3,7,8\}\to \{3\}  $&$\frac{1}{ \sqrt{3}}    $&$\vec  \alpha_1     $  \\ \hline
$           \{4,7,8\}\to \{4\}  $&$-\frac{1}{ \sqrt{3}}
$&$         \{5,7,8\}\to \{5\}  $&$-\frac{1}{ \sqrt{3}}
$&$         \{6,7,8\}\to \{6\}  $&$-\frac{1}{ \sqrt{3}}   $&$\vec  \alpha_4     $  \\ \hline  \hline
$           \{0,1,7\}\to \emptyset  $&$      0
$&$         \{0,2,7\}\to \{2\}  $&$-\frac{1}{\sqrt{2}}
$&$         \{0,3,7\}\to \{3\}  $&$\frac{1} {\sqrt{2}}     $&$\vec  \alpha_2     $  \\ \hline
$           \{0,4,7\}\to \emptyset  $&$      0
$&$         \{0,5,7\}\to \{5\}  $&$\frac{1}{\sqrt{2}}
$&$         \{0,6,7\}\to \{6\}  $&$- \frac{1}{\sqrt{2}}    $&$ \vec \alpha_5     $  \\ \hline\hline
$           \{0,1,8\}\to \{1\}  $&$ - \sqrt{\frac{2}{3}}
$&$         \{0,2,8\}\to \{2\}  $&$\frac{1}{\sqrt{6}}
$&$         \{0,3,8\}\to \{3\}  $&$ \frac{1}{\sqrt{6}}            $&$\vec  \alpha_3     $  \\ \hline
$           \{0,4,8\}\to \{4\}  $&$ \sqrt{\frac{2}{3}}
$&$         \{0,5,8\}\to \{5\}  $&$-\frac{1}{\sqrt{6}}
$&$         \{0,6,8\}\to \{6\}  $&$-\frac{1}{\sqrt{6}}            $&$ \vec \alpha_6     $  \\ \hline  \hline
\end{tabular}
\end{table}
\end{scriptsize}

\begin{eqnarray}
&&\{\vec H_{\alpha},Q_1\}_{S_3}=\vec \alpha_1 Q_1,  \qquad
\{ \vec H_{\alpha},Q_4\}_{S_3}=\vec \alpha_4 Q_4,    \nonumber\\
&&\{\vec H_{\alpha},Q_2\}_{S_3}=\vec \alpha_2 Q_2,  \qquad
 \{\vec H_{\alpha},Q_5\}_{S_3}=\vec \alpha_5 Q_5 \nonumber\\
&&\{\vec H_{\alpha},Q_3\}_{S_3}=\vec \alpha_3 Q_3, \qquad
 \{\vec H_{\alpha},Q_6\}_{S_3}=\vec \alpha_6 Q_6 \nonumber\\
\end{eqnarray}

where $\vec H_{\alpha}=\{  H_1,H_2,H_3\}=\{(Q_7,Q_8), (Q_0,Q_7),(Q_0,Q_8)\}$ and

\begin{eqnarray}
&&\vec \alpha_1=-\vec \alpha_4=\{\frac{1}{ \sqrt{3}}, 0,- \sqrt{\frac{2}{3}} ,\}                         \nonumber\\
&&\vec \alpha_2=-\vec \alpha_5=\{\frac{1}{ \sqrt{3}},-\frac{1}{\sqrt{2}}, \frac{1}{\sqrt{6}}\}
\nonumber\\
&&\vec \alpha_3=-\alpha_6=\{\frac{1}{ \sqrt{3}}, \frac{1}{\sqrt{2}}, \frac{1}{\sqrt{6}}\},\nonumber\\
\end{eqnarray}
where
\begin{eqnarray}
&&< \vec \alpha_i, \vec \alpha_i>=1,  \qquad i=1,2,...,6, \nonumber\\
&&< \vec \alpha_i, \vec \alpha_j>=0, \qquad i,j=1,2,...,6,
\qquad i \neq j.\nonumber\\
\end{eqnarray}

Note, that
\begin{eqnarray}
&&< \vec \alpha_i^0, \vec \alpha_i^0 >=\frac{2}{3},  \qquad i=1,2,...,6,\nonumber\\
&&\vec \alpha_1^0 + \vec \alpha_2^0 + \vec \alpha_3^0 =0,\nonumber\\
&&< \vec \alpha_1^0, \vec \alpha_2^0 >=< \vec \alpha_2^0, \vec \alpha_3^0 >
=< \vec \alpha_3^0, \vec \alpha_1^0 >=-\frac{1}{3}, \nonumber\\
\end{eqnarray}
where
\begin{eqnarray}
\vec \alpha_i^0=\{0,\vec\alpha_{i2}, \vec\alpha_{i3}\},  \qquad i=1,2,...,6,
\end{eqnarray}
are the binary non-zero roots, in which the first components $\vec\alpha_{i1}$, $i=1,...,6$, are equal zero.
Thus, we have got
\begin{eqnarray}
\frac{< \vec \alpha_i, \vec \alpha_i>}
{< \vec \alpha_i^0, \vec \alpha_i^0>}
=\frac{3}{2}
\end{eqnarray}

Note,  that there is only one simple root, since all $\alpha_i$, $i=1,2,3$ or $i=4,5,6$, are related by
usual $Z_2$ transformations, $\vec \alpha_i=-\vec \alpha_{i+3}$, $(i=1,2,3)$,  or by $Z_3$ transformations:
\begin{eqnarray}
\vec \alpha_2=R^{V}(q) \vec  \alpha_1=O(2\pi /3)\vec \alpha_1,\qquad
\alpha_3=R^{V}(q)^2\vec \alpha_1=O(4\pi /3)\vec \alpha_1,
\end{eqnarray}
where
\begin{eqnarray}
R^{V}(q)=O(2\pi /3)=\left(
\begin{array}{ccc}
1 & 0 & 0 \\
0 & -1/2 & \sqrt{3}/2 \\
0 & -\sqrt{3}/2 & -1/2
\end{array}
\right),
\end{eqnarray}
$R^{V}(q^{2})=(R^{V}(q))^{2}$ and $(R^{V}(q))^{3}=R^{V}(q_{0})$

 Now one can unify in second  group  the  other  18  commutations relations:
\begin{tiny}
\begin{table}
\centering \caption{ \it  II-The dual  roots}
\label{COMWeyl} \vspace{.05in}
\begin{tabular}{||c|c||c|c||c|c|c|} \hline
$\{klm\} \rightarrow \{n\} $&$ f_{klm}^n $&$ \{klm\} \rightarrow  \{n\} $&$f_{klm}^n $&$ \{klm\} \rightarrow \{n\}$&$f_{klm}^n  $&$  \vec   \beta_i
$ \\ \hline \hline
$           \{0,1,2\}\to \{6\}  $&$   \frac{1}{\sqrt{3}}
$&$         \{7,1,2\}\to \{6\}  $&$-\frac{\sqrt{2}}{\sqrt{3}}
$&$         \{8,1,2\}\to \{6\}  $&$   \sqrt{2}            $&$  \vec   \beta_1
$  \\ \hline
$           \{0,4,5\}\to \{3\}  $&$ -\frac{1}{\sqrt{3}}
$&$         \{7,4,5\}\to \{3\}  $&$ \frac{\sqrt{2}}{\sqrt{3}}
$&$         \{8,4,5\}\to \{3\}  $&$-\sqrt{2}                $&$  \vec   \beta_4
$  \\ \hline\hline
$           \{0,2,3\}\to \{4\}  $&$   \frac{1}{\sqrt{3}}
$&$         \{7,2,3\}\to \{4\}  $&$\frac{2\sqrt{2}}{\sqrt{3}}
$&$         \{8,2,3\}\to \{\emptyset  \}    $&$  0       $&$  \vec   \beta_2
$ \\ \hline
$           \{0,5,6\}\to \{1\}  $&$-\frac{1}{\sqrt{3}}
$&$         \{7,5,6\}\to \{1\}  $&$- \frac{2\sqrt{2}}{\sqrt{3}}
$&$         \{8,5,6\}\to \{\emptyset  \}    $&$0         $&$   \vec  \beta_5
$ \\ \hline  \hline
$           \{0,3,1\}\to \{5\}  $&$\frac{1}{\sqrt{3}}
$&$         \{7,3,1\}\to \{5\}  $&$-\frac{\sqrt{2}}{\sqrt{3}}
$&$         \{8,3,1\}\to \{5\}  $&$-\sqrt{2}                   $&$   \vec  \beta_3
$ \\ \hline
$           \{0,6,4\}\to \{2\}  $&$-\frac{1}{\sqrt{3}}
$&$         \{7,6,4\}\to \{2\}  $&$ \frac{\sqrt{2}}{\sqrt{3}}
$&$         \{8,6,4\}\to \{2\}  $&$ \sqrt{2}                 $&$  \vec   \beta_6
$ \\ \hline  \hline
\end{tabular}
\end{table}
\end{tiny}

Using the properties of multiplications:
\begin{eqnarray}
&&Q_6=Q_1Q_2, Q_4=Q_2Q_3, Q_5=Q_3Q_1, \nonumber\\
&&Q_3=Q_4Q_5, Q_1=Q_5Q_6, Q_2=Q_6Q_4 ,\nonumber\\
\end{eqnarray}
one can introduce the new systems of the $beta$-roots (see Table II):

\begin{eqnarray}
&&\vec \beta_1=-\vec \beta_4=\{ \frac{1}{\sqrt{3}},-\frac{\sqrt{2}}{\sqrt{3}},\sqrt{2}   \}          \nonumber\\
&&\vec \beta_2=-\vec \beta_5=\{ \frac{1}{\sqrt{3}}, \frac{2\sqrt{2}}{\sqrt{3}},  0\}          \nonumber\\
&&\vec \beta_3=-\vec \beta_6=\{ \frac{1}{\sqrt{3}} ,-\frac{\sqrt{2}}{\sqrt{3}},-\sqrt{2}   \}          \nonumber\\
\end{eqnarray}

\begin{eqnarray}
&& < \vec  \beta_i, \vec \beta_i>=3,  \qquad i=1,2,...,6,\nonumber\\
&&< \vec  \beta_i, \vec \beta_j>=-1, \qquad i,j=1,2,3,   \qquad i\neq j.\nonumber\\
\end{eqnarray}

Note,  that there is also only one simple dual root, since all $\beta_i$, $i=1,2,3$ or $i=4,5,6$, are related by
usual $Z_2$ transformations, $\vec \beta_i=-\vec \beta_{i+3}$, $(i=1,2,3)$,  or by $Z_3$ transformations:
\begin{eqnarray}
\vec \beta_2=R^{V}(q) \vec  \beta_1=O(2\pi /3)\vec \beta_1,\qquad
\beta_3=R^{V}(q)^2\vec \beta_1=O(4\pi /3)\vec \beta_1,
\end{eqnarray}
where
\begin{eqnarray}
R^{V}(q)=O(2\pi /3)=\left(
\begin{array}{ccc}
1 & 0 & 0 \\
0 & -1/2 & \sqrt{3}/2 \\
0 & -\sqrt{3}/2 & -1/2
\end{array}
\right),
\end{eqnarray}
$R^{V}(q^{2})=(R^{V}(q))^{2}$ and $(R^{V}(q))^{3}=R^{V}(q_{0})$

The third group contains itself 27 commutation relations amonth them there are only 9
have the non-zero results:
\begin{table}
\begin{tiny}
\centering
\caption{ \it  III-The  root system of the step operators}
\label{COMWeyl}
\vspace{.05in}
\begin{tabular}{||c|c||c|c||c|c|} \hline
$           \{klm\} \rightarrow \{n\} $&$ f_{klm}^n
$&$         \{klm\} \rightarrow \{n\} $&$ f_{klm}^n
$&$         \{klm\} \rightarrow \{n\} $&$ f_{klm}^n
$  \\ \hline \hline
$           \{1,4,0\}\to \{0,7,8\}    $&$\{0,0,\frac{\sqrt{2}}{\sqrt{3}}\}
$&$         \{1,5,0\}\to \emptyset    $&$\{0,0,0 \}
$&$         \{1,6,0\}\to \emptyset    $&$\{0,0,0 \}
$  \\ \hline
$           \{1,4,7\}\to \{0,7,8\}    $&$\{0,0,\frac{1}{ \sqrt{3}}\}
$&$         \{1,5,7\}\to \emptyset    $&$\{0,0,0 \}
$&$         \{1,6,7\}\to \emptyset    $&$\{0,0,0 \}
$  \\ \hline
$           \{1,4,8\}\to \{0,7,8\}    $&$\{\sqrt{6},\sqrt{3},  0 \}
$&$         \{1,5,8\}\to \emptyset    $&$\{0,0,0 \}
$&$         \{1,6,8\}\to \emptyset    $&$\{0,0,0 \}
$  \\ \hline \hline
$           \{2,4,0\}\to \emptyset    $&$ \{0,0,0 \}
$&$         \{2,5,0\}\to \{0,7,8\}    $&$ \{0,\frac{1}{\sqrt{2}},-\frac{1}{\sqrt{6}}\}
$&$         \{2,6,0\}\to \emptyset    $&$ \{0,0,0 \}
$  \\ \hline
$           \{2,4,7\}\to \emptyset    $&$ \{0,0,0 \}
$&$         \{2,5,7\}\to \{0,7,8\}    $&$ \{\frac{3}{\sqrt{2}},-1,-\frac{2}{\sqrt{3}}\}
$&$         \{2,6,7\}\to \emptyset    $&$ \{0,0,0 \}
$  \\ \hline
$           \{2,4,8\}\to \emptyset    $&$ \{0,0,0 \}
$&$         \{2,5,8\}\to \{0,7,8\}    $&$ \{-\frac{\sqrt{3}}{\sqrt{2}},0,1\}
$&$         \{2,6,8\}\to \emptyset    $&$ \{0,0,0 \}
$  \\ \hline  \hline
$           \{3,4,0\}\to \emptyset    $&$ \{0,0,0 \}
$&$         \{3,5,0\}\to \emptyset    $&$ \{0,0,0 \}
$&$         \{3,6,0\}\to \{0,7,8\}    $&$ \{0,-\frac{1}{\sqrt{2}},-\frac{1}{\sqrt{6}}\}
$  \\ \hline
$           \{3,4,7\}\to \emptyset     $&$ \{0,0,0 \}
$&$         \{3,5,7\}\to \emptyset    $&$ \{0,0,0 \}
$&$         \{3,6,7\}\to \{0,7,8\}    $&$ \{-\frac{3}{\sqrt{2}},1,-\frac{2}{\sqrt{3}}\}
$  \\ \hline
$           \{3,4,8\}\to \emptyset     $&$ \{0,0,0 \}
$&$         \{3,5,8\}\to \emptyset    $&$ \{0,0,0 \}
$&$         \{3,6,8\}\to \{0,7,8\}    $&$ \left\{-\sqrt{\frac{3}{2}},0,-1\right\}
$  \\ \hline  \hline
\end{tabular}
\end{tiny}
\end{table}

The fourth group has also the 18 commutations relations:
\begin{table}
\centering
\caption{ \it  IV-The  root system of the step operators}
\label{COMWeyl}
\vspace{.05in}
\begin{tabular}{||c|c||c|c||c|c|} \hline
$           \{klm\} \rightarrow \{n\} $&$ f_{klm}^n
$&$         \{klm\} \rightarrow  \{n\}$&$ f_{klm}^n
$&$         \{klm\} \rightarrow \{n\} $&$ f_{klm}^n
$  \\ \hline \hline
$           \{1,2,4\}\to \{2\}            $&$\{1\}
$&$         \{1,2,5\}\to \{1\}            $&$\{1\}
$&$         \{1,2,6\}\to \{\emptyset\}    $&$\{0 \}
$  \\ \hline
$           \{1,3,4\}\to \{3\}            $&$\{-1\}
$&$         \{1,3,5\}\to \{\emptyset \}   $&$\{0\}
$&$         \{1,3,6\}\to \{1\}            $&$\{-1\}
$  \\ \hline
$           \{2,3,4\}\to \{\emptyset\}    $&$\{0\}
$&$         \{2,3,5\}\to \{3\}            $&$\{1\}
$&$         \{2,3,6\}\to \{2\}            $&$\{1\}
$  \\ \hline \hline
$           \{1,4,5\}\to \{5\}            $&$\{-1\}
$&$         \{2,4,5\}\to \{4\}            $&$\{-1\}
$&$         \{3,4,5\}\to \{\emptyset\}    $&$\{0\}
$  \\ \hline
$           \{1,4,6\}\to \{6\}            $&$\{1\}
$&$         \{2,4,6\}\to \{\emptyset\}    $&$\{0\}
$&$         \{3,4,6\}\to \{4\}            $&$\{1\}
$  \\ \hline
$           \{1,5,6\}\to \{\emptyset\}    $&$\{0\}
$&$         \{2,5,6\}\to \{6\}            $&$\{-1\}
$&$         \{3,5,6\}\to \{5\}            $&$\{-1\}
$  \\ \hline
\end{tabular}
\end{table}

The last, 5-th , group has only two but very important commutation relations:

\begin{table}
\centering
\caption{ \it  V-The  root system of the step operators}
\label{COMWeyl}
\vspace{.05in}
\begin{tabular}{||c|c||c|c|} \hline
$           \{klm\} \rightarrow \{n\} $&$ f_{klm}^n
$&$         \{klm\} \rightarrow  \{n\}$&$ f_{klm}^n
$  \\ \hline \hline
$           \{1,2,3\}\to \{0\}            $&$\{\sqrt{3}\}
$&$         \{4,5,6\}\to \{0\}            $&$\{-\sqrt{3}\}
$  \\ \hline
\end{tabular}
\end{table}

\section{${C}_N$- Clifford algebra}

We begin with a $V$, a finite-dimensional  vector space over the fields,
$\Lambda={ R},{ C} $ or  $\Lambda={ TC}$.

We introduce the tensor algebra $T(V)=\oplus_{n \geq 0}\otimes^n V$,
with ${\otimes}^0 V=\Lambda$.

 The product in $T(V)$ one can define as follows:
$v_1\otimes ...\otimes v_p\in V^{\otimes p}$  and
$u_1\otimes ...\otimes u_q \in V^{\otimes q}$, then their product is
$v_1\otimes ...\otimes v_p\otimes u_1\otimes ...\otimes u_q \in V^{\otimes (p+q)}$.
For example, if $V$ has a basis $\{x,y\}$, then $T(V)$ has a basis $\{1,x,y,xy,yx,x^2,y^2,
x^2y,y^2x,x^2y^2,....\}$.
Suppose now we introduce into $V$ a trilinear form $(...,...,...)$.
Let $J=<v\otimes v \otimes v-(v,v,v)\cdot 1| v \in V>$ an ideal in $T(V)$ and put
\begin{eqnarray}
TCl(V)=T(V)/J,
\end{eqnarray}
the Clifford algebra over $V$ with trilinear form
$(...,...,...)$.

To generalize the binary Clifford algebra one can introduce the  following
generators $q_1,q_2,...,q_n$ and relations:
\begin{eqnarray}
q_k^3=1
\end{eqnarray}
and
\begin{eqnarray}
q_kq_l=jq_lq_k, \qquad q_lq_k=j^2q_kq_l,\qquad n\geq l>k\geq1,
\end{eqnarray}
where $j=\exp{(2\pi /3)}$.
One can immeadiately find two types  of the $S_3$ identities.
The first type of such identities are:
\begin{eqnarray}
&&q_k q_l q_k +     q_k^2 q_l +    q_l q_k^2 = (j+1+j^2) q_k^2 q_l=0, \nonumber\\
&&q_k q_l q_k +  j^2q_k^2 q_l + j  q_l q_k^2 = (j+j^2+1) q_k^2 q_l=0, \nonumber\\
&&q_k q_l q_k +  j  q_k^2 q_l + j^2q_l q_k^2 = (3j)      q_k^2 q_l,   \nonumber\\
\end{eqnarray}
or

\begin{eqnarray}
&&q_l q_k q_l +     q_1^2 q_k +    q_k q_l^2 = (j^2+j+1) q_kq_l^2 =0, \nonumber\\
&&q_l q_k q_l + j^2 q_1^2 q_k +  j q_k q_l^2 = (j^2+1+j) q_kq_l^2 =0, \nonumber\\
&&q_l q_k q_l + j   q_1^2 q_k + j^2q_k q_l^2 = (3j^2)    q_kq_l^2 , \nonumber\\
\end{eqnarray}

The second type of the identities relate to the triple product of the generators
with all different indexes,
for example , one can take take $n \geq m >l>k\geq 1 $. Then one can easily get :
\begin{eqnarray}
&& (q_k q_l q_m +     q_l q_m q_k +    q_m q_k q_l)
 + (q_m q_l q_k +     q_l q_k q_m +    q_k q_m q_l) \nonumber\\
&=&\biggl (1+j^2+j^2)+(1 +j+j)\biggl )q_kq_lq_m
 = \biggl ((j^2-j)+(j-j^2)    \biggl )q_kq_lq_m     \nonumber\\
&=&\biggl (1+j+j)+(1 +j^2+j^2)\biggl )q_mq_lq_k
 = \biggl ((j-j^2)+(j^2-j)    \biggl )q_mq_lq_k     \nonumber\\
&=&0.                                               \nonumber\\
\end{eqnarray}

From these two types of the identities one can see, that $S_{3+}$-symmetric
sum
\begin{eqnarray}
\sum_{S_{3+}}=q_k q_l q_m +     q_l q_m q_k +    q_m q_k q_l
 + q_m q_l q_k +     q_l q_k q_m +    q_k q_m q_l=\{q_kq_lq_m\}
\end{eqnarray}

is not equal zero just in one case, when all indexes, $k,l,m$ are equal,
{\it i.e.}:
\begin{eqnarray}
\sum_{S_{3+}}(q_k q_l q_m )= q_k q_l q_m +     q_l q_m q_k +    q_m q_k q_l
 + q_m q_l q_k + q_l q_k q_m +    q_k q_m q_l=6 \delta_{klm}.
\end{eqnarray}

\begin{eqnarray}
\sum_{k=1}^{k=n}\sum_{l=1}^{l=n}\sum_{m=1}^{m=n}(q_kq_lq_m)=
\sum_{S_{3+}}(q_k q_l q_m )= (q_k q_l q_m +     q_l q_m q_k +    q_m q_k q_l
 + q_m q_l q_k +     q_l q_k q_m +    q_k q_m q_l)=n \delta_{klm}.
\end{eqnarray}

$TCl(V)$ is a $Z_3$-graded algebra.
We put
\begin{eqnarray}
T(V)_0=\oplus_{n=3k}   V^{\otimes n}, \qquad
T(V)_1=\oplus_{n=3k+1} V^{\otimes n}, \qquad
T(V)_2=\oplus_{n=3k+2} V^{\otimes n}.
\end{eqnarray}
Also, one can see
\begin{eqnarray}
TCl(V)_0=TCl(V)_0 \oplus TCl(V)_1 \oplus TCl(V)_2, \qquad TCl(V)_k=T(V)_k/J_k.
\end{eqnarray}
\begin{eqnarray}
J_k=J\bigcap T(V)_k.
\end{eqnarray}

If $dim_{\Lambda} V=n$ and $\{q_1,...,q_n\}$ is an ortghonal basis for $V$ with
$(q_k,q_l,q_m)=\lambda_k \delta_{k,l,m}$, then the dimension
$dim_{\Lambda}TCl(V)=3^n$ and $\{\prod q_k^{l_k}\}$ is a basis where $l_k$
is $0$,$1$,or $2$.

\begin{eqnarray}
\begin{array}{|c|c|c|c|c|c|c|}
n -gen   &   1  &  2 &  3   & 4       &  5       &  6  \\ \hline
TCl_0    &   1  & 1+2& 1+7+1& 1+16+10 &1+30+45+5 & 1+50+141+50+1 \\
TCl_1    &   1  & 2+1& 3 +6 & 4+19+4  &5+45+30+1 &  -               \\
TCl_2    &   1  &  3 & 6 + 3& 1+16+1  &15+51+15  &  -                    \\\hline
 \Sigma  &   3  & 3\times 3= 9 & 9\times 3  &  81     & 81\times 3=243
 & 243 \times 3=729 \\\hline
 \end{array}
\end{eqnarray}

\begin{eqnarray}
\begin{array}{|c|c|c|}\hline
 0.    &     1                                                                            &1\\ \hline
 1.    &< q_1,q_2,q_3,q_4>                                                                &4  \\ \hline
 2.    &<q_1^2,q_2^2,q_3^2,q_4^2>                                                         & - \\
       & <q_1q_2,q_1q_3,q_1q_4,q_2q_3,q_2q_4,q_3q_4>                                      &10\\   \hline
 3.    &<q_1^2q_2,q_1^2q_3,q_1^2q_4,q_2^2q_1, q_2^2q_3,q_2^2q_4,                          &  - \\
       & q_3^2q_1,q_3^2q_2,q_3^2q_4,q_4^2q_1,q_4^2q_2,q_4^2q_3>                           & -  \\
       &<q_1q_2q_3,q_1q_2q_4,q_1q_3q_4, q_2q_3q_4>                                        &16\\ \hline
 4.    &<q_1^2q_2^2,q_1^2q_3^2,q_1^2q_4^2,q_2^2q_3^2,q_2^2q_4^2,q_3^2q_4^2>               & -\\
       & <q_1^2q_2q_3,  q_1^2q_2q_4,q_1^2q_2q_3,q_2^2q_1q_3,q_2^2q_1q_4,q_2^2q_3q_4,      &   - \\
       & q_2^2q_1q_3,q_2^2q_1q_4,q_2^2q_3q_4, q_2^2q_1q_3,q_2^2q_1q_4,q_2^2q_3q_4         & -\\
       & q_3^2q_1q_2,q_3^2q_1q_4,q_3^2q_2q_4, q_4^2q_1q_2,q_4^2q_1q_3,q_4^2q_2q_3>        &- \\
       &  <q_1q_2q_3q_4>                                                                  & 19 \\ \hline
 5.    &<q_1^2q_2^2q_3, q_1^2q_2^2q_4, q_1^2q_3^2q_2,q_1^2q_3^2q_4, q_1^2q_4^2q_2,q_1^2q_4^2q_3, & - \\
       & q_2^2q_3^2q_1, q_2^2q_3^2q_4, q_2^2q_4^2q_1,q_2^2q_4^2q_3, q_3^2q_4^2q_1, q_3^2q_4^2q_2> &- \\
       & <q_1^2q_2q_3q_4,q_2^2q_1q_3q_4,q_3^2q_1q_2q_4,q_4^2q_1q_2q_3>                    &16\\ \hline
 6.    &<q_1^2q_2^2q_3^2,q_1^2q_2^2q_4^2,q_1^2q_3^2q_4^2,q_2^2q_3^2q_4^2>                 & - \\
       & <q_1^2q_2^2q_3q_4, q_1^2q_3^2q_2q_4,q_1^2q_4^2q_2q_3,
          q_2^2q_3^2q_1q_4,q_2^2q_4^2q_1q_3,q_3^2q_4^2q_1q_2>                             & 10\\ \hline
 7.    &<q_1^2q_2^2q_3^2q_4,q_2^2q_3^2q_4^2q_1,q_3^2q_4^2q_1^2q_2,q_4^2q_1^2q_2^2q_3>     & 4 \\ \hline
 8     &< q_1^2q_2^2q_3^2q_4^2> &  1 \\ \hline
\end{array}
\end{eqnarray}


Acknowledments.
We are very grateful   to Igor Ajinenko, Luis Alvarez-Gaume, Ignatios Antoniadis, Genevieve Belanger, Nikolai Boudanov,
 Tatjana Faberge, M. Vittoria Garzelli,
 Nanie Perrin, Alexander Poukhov, for very nice  support.
Some important results we  have got from very  nice discussions with
Alexey Dubrovskiy, John Ellis, Lev Lipatov, Richard Kerner,
Andrey Koulikov, Michel Rausch de Traunberg,
Robert Yamaleev.
Thank them very much.


\begin{thebibliography}{99}

\bibitem{LRV}
L. N. Lipatov, M. Rausch de Traunbenberg,  G. Volkov,
{\it On the ternary complex analysis and its applications}
J. Math. Phys. {\bf 49} 013502 (2008)\\

\bibitem{LV}
{ On the  complexifications \\
of the Euclidean $R^n$ spaces
and the n-dimensional generalization of Pithagore theorem}\\
in preparation.

\bibitem{Volkov1}
{The GUT of the light and complexification of all $R^n$ Euclidean spaces}
Arxiv, General Physics.


\bibitem{Hamilton}
R.W. Hamilton, {\it On Quaternions}, Proceedings of the Royal
Irish Academy, Nov.11 (1844), {\bf v.3} (1847), 1-16.\\



\bibitem{FTY}
N. Fleury, M. Rausch de Traunbenberg et R.M. Yamaleev, {\it J.
Math. Anal. and Appl.}, {\bf 180}, (1993), 431.\\


\bibitem{FTY2}
N. Fleury, M. Rausch de Traunbenberg et R.M. Yamaleev, {\it J.
Math. Anal. and Appl.}, {\bf 191}, (1995), 118. \\




\bibitem{Kern}
R. Kerner, {\it Ternary algebraic structures and their
applications in physics}, Proceedings of the Conference ICGTMP
"Group-23", sis Dubna,2000, Russia,math-ph/0011023. \\



\bibitem{DV}
A.Dubrovskiy and G.Volkov,
{\it Ternary numbers and algebras. Reflexive numbers and Berger graphs }
Adv.Appl.CliffordAlgebras17:159-181,2007\\
archiv:hep-th/0608073, (2006).\\





\bibitem{Berger}
M. Berger, {\it Sur les groupes d'holonomie homog\`ene des
vari\'et\'es \'a connexion affine et des vari\'et\'es
riemanniennes},
Bull. Soc. Math. France {\bf 83} (1955) 279--310.\\



\bibitem{AENV1}
F. Anselmo, J. Ellis, D. Nanopoulos, G. Volkov, {\it Towards an
algebraic classification of Calabi-Yau Manifolds: Study of K3
spaces}, Phys. Part. Nucl. {\bf 32}(2001) 318-375; Fiz. Elem.
Chast. Atom. Yadra {\bf 32} (2001) 605-698.\\

\bibitem{AENV2}
 F. Anselmo, J. Ellis, D.V. Nanopoulos, G. Volkov\\
 {\ it Universal Calabi-Yau Algebra: Towards an Unification of Complex Geometry}\\
Int.J.Mod.Phys. A18 (2003) 5541-5612 \\
archiv:hep-th/0207188.\\



\bibitem{V}
G. Volkov, {\it Hunting for the New Symmetries in Calabi-Yau
Jungles}, {Int.J. Mod. Phys} {\bf A19} (2004) 4835-4860,
hep-th/0402042.\\






\bibitem{Traubenberg}
M. Raush de Traubenberg, {\it Alg{\`e}bres de Clifford
Supersym/'etrie et Sym/'etrie $Z_n$. Applications en Th/'eorie des
Champs}, LPT 97-02, 1997.\\




\end{thebibliography}
\end{document}